\documentclass[%
 reprint,
showpacs,preprintnumbers,
 amsmath,amssymb,
 aps,pra
]{revtex4-1}

\usepackage{graphicx}% Include figure files
\usepackage{dcolumn}% Align table columns on decimal point
\usepackage{bm}% bold math
\newcommand{\ket}[1]{\ensuremath{| #1 \rangle}}
\newcommand{\bra}[1]{\ensuremath{\langle #1 |}}
\newcommand{\ave}[1]{\ensuremath{\langle #1 \rangle}}
\newcommand{\e}{\mathrm{e}}

\begin{document}

\title{Quantum state reconstruction of spectral field modes: homodyne and resonator detection schemes}

\author{F. A. S. Barbosa$^1$, A. S. Coelho$^1$, K. N. Cassemiro$^2$, P. Nussenzveig$^1$, C. Fabre$^3$, A. S. Villar$^2$, and M. Martinelli$^1$}

\email{mmartine@if.usp.br}

\affiliation{$^1$ Instituto de F\'\i sica, Universidade de S\~ao Paulo, P.O. Box 66318, 05315-970 S\~ao Paulo, Brazil. \\
$^2$ Departamento de F\'\i{}sica, Universidade Federal de Pernambuco, 50670-901 Recife, Brazil. \\
$^3$ Laboratoire Kastler Brossel, Universit\'e Pierre et Marie Curie -- Paris 6, 75252 Paris, France. }

\begin{abstract}
We revisit the problem of quantum state reconstruction of light beams from the photocurrent quantum noise. 
As is well-known, but often overlooked, two longitudinal field modes contribute to each spectral component 
of the photocurrent (sideband modes). We show that spectral homodyne detection is intrinsically incapable 
of providing all the information needed for the full reconstruction of the two-mode spectral quantum state.  
Such a limitation is overcome by the technique of resonator detection. A detailed theoretical description and comparison 
of both methods is presented, as well as an experiment to measure the six-mode quantum state of 
pump-signal-idler beams of an optical parametric oscillator above the oscillation threshold. 
\end{abstract}

 % final suggestion

\maketitle

Quantum optics employing continuous variables of the electromagnetic field is a mature and 
well-developed subject, with applications ranging from high resolution measurements~\cite{LIGO}, 
to manipulation and storage of quantum information~\cite{teleport1,teleport2,storeQI}, and 
quantum metrology~\cite{morganmitchell}. Among its advantages are the use of techniques 
adapted from the classical communications community, which employ the spectral analysis of 
light~\cite{naturephotrev}. Quantum features that play a role in these applications include quadrature 
squeezing~\cite{squeezing}, quantum correlations~\cite{twin beams} and entanglement~\cite{entangledkimble}.

In order to harness the advantages offered by quantum properties of light to improve high resolution 
measurements or quantum information protocols, it is often necessary to obtain full knowledge of 
the system's quantum-mechanical state. Techniques for complete quantum-state characterization have been 
a part of the quantum optics toolbox for 20 years~\cite{raymer,raymerrmp}. However, when combining 
these techniques with the spectral analysis of measured signals~\cite{yurkemedidasqzPRA85}, care must be exercised: 
it has been known for a long time that two (sideband) modes must be considered when measuring quantum noise 
(and correlation) spectra of a single beam of light. In many situations, an effective single-mode 
description can be applied but this is not always true. 

In a previous paper~\cite{prltobe}, we show experimentally that indeed two different light 
states could lead to the same homodyne detection signals, whereas they could be unambiguously 
discriminated by resonator detection. In the present paper, our purpose is to give a detailed 
and consistent description of spectral reconstruction of quantum states of light. For the sake of 
completeness, in part of the paper we review concepts that are already known (although 
sometimes neglected). This helps make clear the shortcomings of the most widely used detection 
technique, (spectral) homodyne detection (HD), as well as the demonstration that an 
alternative technique, resonator detection (RD)~\cite{levenson,shelby}, does not suffer from the same limitations. 

Information about the quantum state is retrieved from photodetection, which yields a 
photocurrent continuously varying in time. Interferometric techniques, usually involving a reference 
field (a Local Oscillator - LO), enable the acquisition of phase sensitive information, thus allowing 
the measurement of field quadratures. In HD, a weak signal field is combined with a strong 
local oscillator (assumed to be well described by a coherent state) on a beam splitter with 
balanced reflection and transmission. The two outputs are detected and their photocurrents 
combined. The temporal behavior of the photocurrent is determined by the beating of the 
LO (carrier) mode with other modes slightly detuned by positive and negative amounts. 
When directly analyzing the photocurrent in the temporal domain, the effect of these neighboring 
modes is integrated within a bandwidth determined by the temporal resolution of the measurement. 
This constitutes a single ``temporal'' mode and provides an adequate description for measuring a beam 
of light~\cite{optcommunLvovsky}. On the other hand, the spectral analysis aims at resolving individual 
spectral modes, i.e. with a given frequency. This requires beating the photocurrent with a radio 
frequency (rf) reference field. 
The beatnote signal at a given analysis frequency comprises both sidebands 
symmetrically detuned with respect to the LO frequency (upper and lower sidebands), without 
distinguishing between them. Thus, the full determination of the quantum state of a beam of light at a 
given frequency requires characterizing each of the two modes, as well as the correlations 
between them. 

In many situations, a change of basis to symmetric ($\mathcal{S}$) and anti-symmetric ($\mathcal{A}$) combinations of the 
upper and lower sidebands results in an effective single-mode description. State reconstruction 
is however limited to situations in which there is no coupling between the $\mathcal{S}$ and $\mathcal{A}$ modes. An 
example is the measurement of single-beam squeezing at a given frequency: it has been known 
for long and it was experimentally demonstrated that squeezing of the 
$\mathcal{S}$ (or $\mathcal{A}$) mode corresponds to entanglement between the upper and lower sidebands~\cite{zhangPRA2003,huntingtonPRA05}. We show 
below that HD is intrinsically ``blind'' to correlations between $\mathcal{S}$ and $\mathcal{A}$ modes of a single beam. 
Physically, this is a result of the perfect symmetry between upper and lower sidebands in the 
detection process. In contrast, in RD the field modes interact with an empty optical resonator prior to photodetection. 
The upper and lower sidebands undergo different phase shifts and, especially, different attenuations 
when reflected from the empty cavity as a function of its detuning. This constitutes a previously unknown 
and unrecognized advantage of RD when compared to HD. A complete measurement of all second 
order moments suffices to fully characterize a Gaussian quantum state. This is possible with RD but 
unattainable with HD.

In this paper, after defining a notation for the covariance matrix, treated as a complete representation 
of any Gaussian state (section~\ref{sec:quantumstatereconstruction}), we review the description of photocurrent as a quantum measurement (sec.~\ref{sec:photodetection}).
We highlight  the measurement operators associated with homodyne detection and its limitations regarding the reconstruction of quantum states (sec.~\ref{sec:homodynedetection}). Resonator detection is similarly examined afterwards (sec.~\ref{sec:resonatordetection}). The main result here is the determination of which two-beam correlations remain `hidden' to spectral homodyne detection, and the demonstration of the complete accessibility of the covariance matrix with resonator detection. 

The precise extent to which both techniques differ in a realistic experimental situation is discussed in section~\ref{sec:phasemixingregime}. In most experiments, the electronic Local Oscillator phase is not actively locked to the optical Local Oscillator phase. We discuss the changes to measurement operators and general limitations to the reconstruction of quantum states when performing the spectral analysis of the photocurrent without good phase reference, owing to the optical phase diffusion. 
By extending the treatment to more beams of light (sec.~\ref{sec:generalization}), resonator detection is shown to provide complete state reconstruction of any multimode Gaussian state of spectral modes.
In section~\ref{subsec:efectivesinglemode}, we make the connection between the quantum formalism here utilized and the semi-classical formalism commonly employed in the description of quantum noise. 
Finally, we present experimental results employing resonator detection to show the existence of `hidden' correlations among the three beams (pump, signal, and idler) produced by the OPO (sec.~\ref{sec:experimental}) . The six mode covariance matrix of the measured system is then presented.
We offer our concluding remarks in Section~\ref{sec:conclusao}.

\section{Gaussian quantum states and the covariance matrix}
\label{sec:quantumstatereconstruction}

The class of Gaussian quantum states is particularly important to describe experiments in quantum optics in the continuous variables (CV) domain. Such states are characterized by the observation of Gaussian probability distributions in measurements of quadrature operators (Gaussian Wigner functions). 

For one beam of light, a single longitudinal mode with optical frequency $\omega$ is represented by the amplitude $\hat p_\omega$ and phase $\hat q_\omega$ quadrature observables, satisfying commutation relations $[\hat p_\omega,\hat q_{\omega'}]=2i\delta(\omega-\omega')$. In terms of photon annihilation $\hat a_\omega$ and creation $\hat a^\dag_\omega$ operators, satisfying  $[\hat a_\omega,\hat a^\dag_{\omega'}]=\delta(\omega-\omega')$, they relate as $\hat a_\omega=(\hat p_\omega+i\hat q_\omega)/2$.

Ordering the relevant quadrature operators in a column vector $\vec X = (\hat p_{\omega}\; \hat q_{\omega} \;\hat p'_{\omega'} \;\hat q'_{\omega'} \dots)^T$, the symmetric covariance matrix is defined as 
\begin{equation}
\label{Vmatrix}
\mathbf{V} = \frac12 (\ave{\vec X \cdot \vec{X}^T}+\ave{\vec {X}^T \cdot \vec X}),
\end{equation}
where the average is performed over the quantum state describing the whole quantum field.
Diagonal elements of $\mathbf{V}$ represent variances of single-mode quadrature operators, denoted as e.g. $\Delta^2 \hat p_\omega \equiv \ave{\hat p_\omega\hat p_\omega}$. Off-diagonal elements are correlations between different quadratures operators, such as in e.g. $C(\hat p_\omega \hat p'_{\omega'})\equiv\ave{\hat p_\omega \hat p'_{\omega'}}$.

The covariance matrix completely accounts for the quantum noise of the Gaussian state. For instance, a general two-mode covariance matrix reads as 
\begin{equation}
\mathbf{V}=\left(
\begin{array}{cccc}
\Delta^2 \hat p_{\omega} & C(\hat p_{\omega}\hat q_{\omega}) & C(\hat p_{\omega}\hat p'_{\omega'}) & C(\hat p_{\omega}\hat q'_{\omega'}) \\
 & \Delta^2 \hat q_{\omega} & C(\hat p'_{\omega'}\hat q_{\omega}) & C(\hat q_{\omega}\hat q'_{\omega'})\\
 &  & \Delta^2 \hat p'_{\omega'} & C(\hat p'_{\omega'}\hat q'_{\omega'})\\
 &  & & \Delta^2 \hat q'_{\omega'}
%\Delta^2 \hat p_{\omega} & C_{\hat p_{\omega}\hat q_{\omega}} & C_{\hat p_{\omega}\hat p'_{\omega'}} & C_{\hat p_{\omega}\hat q'_{\omega'}} \\
% & \Delta^2 \hat q_{\omega} & C_{\hat p'_{\omega'}\hat q_{\omega}} & C_{\hat q_{\omega}\hat q'_{\omega'}}\\
% &  & \Delta^2 \hat p'_{\omega'} & C_{\hat p'_{\omega'}\hat q'_{\omega'}}\\
% &  & & \Delta^2 \hat q'_{\omega'}
%\Delta^2 \hat p_{\omega} & C_{\hat p_{\omega}\hat q_{\omega}} & C_{\hat p_{\omega}\hat p'_{\omega'}} & C_{\hat p_{\omega}\hat q'_{\omega'}} \\
% & \Delta^2 \hat q_{\omega} & C_{\hat p'_{\omega'}\hat q_{\omega}} & C_{\hat q_{\omega}\hat q'_{\omega'}}\\
% &  & \Delta^2 \hat p'_{\omega'} & C_{\hat p'_{\omega'}\hat q'_{\omega'}}\\
% &  & & \Delta^2 \hat q'_{\omega'}
\end{array}
\right),
\label{eq:V2mode}
\end{equation}
where repetitive entries have been omitted (since $\mathbf{V}=\mathbf{V}^T$). For a general $n$-mode matrix, $n(2n+1)$ second-order moments fully determine the Gaussian state.

\section{Photodetection}
\label{sec:photodetection}

Photodetectors generate a time-dependent photocurrent $I(t)$ that gives information about the intensity of the incident light beam. In the CV regime, it is a continuous signal deprived of quantum jumps associated with individual quanta of light. In the case of unit quantum efficiency they measure directly the quantum observable $\hat I (t)$ given by~\cite{glauber}
\begin{equation}
\hat I(t) = \hat E^{-}(t)\hat E^{+}(t),
\label{eq:fotocorrente}
\end{equation}
where $\hat E^{\pm}(t)$ are the positive and negative frequency parts of the electric field operator, $\hat E(t)=\hat E^{+}(t)+\hat E^{-}(t)$, written in the case of a narrowband light source, and within a multiplicative factor, as
\begin{equation}
\hat E^+(t) = \int d\omega\, \e^{-i\omega t} \,\hat a_\omega\;, \quad \vec E^-(t)=\left(\vec E^+(\vec r)\right)^\dag,
\label{eq:emaisemenos}
\end{equation}
where the integration limits enclose a frequency interval compatible with the photodetector bandwidth around the optical frequency $\omega_0$ of a bright auxiliary field, the local oscillator (LO). 
Quantum noise results from the `amplification' of quantum fluctuations originating from modes in the frequency vicinity of the LO. 
The LO field must possess a well defined phase relation with respect to the quantum state $\ket{\Psi}$ of remaining modes, and is hence effectively described as a coherent state $\ket{\alpha_{\omega_0}}$, where $\alpha_{\omega_0}=|\alpha|\exp(i\varphi)$ denotes its amplitude and phase. This discussion is easily generalized to account for quantum states represented by density operators.
 
With this general experimental arrangement, valid for the two measurement techniques we analyze in this paper (homodyne and resonator detection schemes), the field quantum state just prior to detection is $\ket{\alpha_{\omega_0}}\ket{\Psi}$.  The quantum state average of Eq.~(\ref{eq:fotocorrente}), together with Eq.~(\ref{eq:emaisemenos}), yields the photocurrent 
\begin{align}
\label{eq:photocu}
I(t) &\propto \int\!\!d\omega\!\!\int\!\!d\omega'\,\e^{i(\omega-\omega') t}\bra{\alpha_{\omega_0}}\bra{\Psi}\hat a^\dag_\omega\,\hat a_{\omega'}\ket{\alpha_{\omega_0}}\ket{\Psi}\nonumber\\
&\approx|\alpha|^2+ |\alpha| \bra{\Psi}\left(\e^{-i\varphi}\hat a(t)+ \e^{i\varphi}\hat a^\dag(t)\right)\ket{\Psi},
\end{align}
where small contributions have been disregarded.

The state-dependent term represents quantum fluctuations of the photocurrent. The operator inside brackets $\delta \hat I(t)=\e^{-i\varphi}\hat a(t)+ \e^{i\varphi}\hat a^\dag(t)$  is the measurement operator, which includes new annihilation and creation operators defined as
\begin{equation}
\label{eq:defat}
\hat a(t)=\int_0^{'\infty} d\omega\,\e^{-i(\omega-\omega_0) t}\,\hat a_{\omega}, \quad \hat a^\dag(t)=[\hat a(t)]^\dag,
\end{equation}
where the integral in $\omega$ must exclude mode $\omega_0$ (a fact denoted by the prime). 

Finally, we note that the response time of a realistic photodetector will necessarily impose the temporal integration of Eq.~(\ref{eq:photocu}), defining the spectral shape of measured mode $\hat a(t)$ in Eq.~(\ref{eq:defat})~\cite{yurkewidebandPRA85}. In temporal homodyne detection, that would define a single-mode field operator (delocalized in frequency or, equivalently, a propagating mode), to which the measurement would correspond~\cite{optcommunLvovsky}. We focus, however, on the spectral analysis of the photocurrent, which we describe next.

\subsection{Photocurrent observable in the spectral domain}
\label{subsec:spectralphotocurrent}

In this paper, we focus on techniques to extract information about the quantum state of light in \textit{spectral modes}. We perform the spectral analysis of the photocurrent to obtain the noise power at a single Fourier frequency $\Omega$, usually in the MHz range~\cite{yurkemedidasqzPRA85}. Low frequency technical noise from multiple sources can then be avoided in the quantum analysis.

The photocurrent fluctuation given by Eq.~(\ref{eq:photocu}) can be described in frequency domain by Fourier transform as 
\begin{equation}
\hat I_{\Omega}= \int\delta\hat I(t)\,\mathrm{e}^{i\Omega t}dt,
\end{equation}
where the integration limits are determined by detection bandwidth. It is easy to show that the spectral component of the photocurrent is a complex quantity associated with the non-Hermitian operator
\begin{equation}
\label{eq:Iomeganonhermitian}
\hat I_\Omega= \mathrm{e}^{-i\varphi}\hat a_u+\mathrm{e}^{i\varphi}\hat a^\dag_{\ell},
\end{equation}
where $\hat a_u$ and $\hat a_{\ell}$ are the annihilation operators of the upper and lower sideband modes at frequencies $\omega_0 + \Omega$ and $\omega_0 - \Omega$, respectively. Therefore, spectral analysis \textit{necessarily implies a two-mode detection scheme}. We note that $\hat I^\dag_\Omega = \hat I_{-\Omega}$. 

The operators $\hat I_\Omega$ are written in terms of quantum mechanical observables $\hat I_\mathrm{cos}$ and $\hat I_\mathrm{sin}$ representing the photocurrent electronic quadratures as $\hat I_\Omega= (\hat I_\mathrm{cos}+ i \hat I_\mathrm{sin})/\sqrt{2}$, where
\begin{align}
\hat I_\mathrm{cos}&{\displaystyle =\cos\varphi\,\frac{\hat p_{u}+\hat p_{\ell}}{\sqrt{2}}+\sin\varphi\,\frac{\hat q_{u}+\hat q_{\ell}}{\sqrt{2}}},\vspace{5pt}\\
\hat I_\mathrm{sin}&={\displaystyle \cos\varphi\,\frac{\hat q_{u}-\hat q_{\ell}}{\sqrt{2}}-\sin\varphi\,\frac{\hat p_{u}-\hat p_{\ell}}{\sqrt{2}}}.
\label{harmoniccomp}
\end{align}
These measurement operators are associated with field modes that are symmetric and anti-symmetric combinations of sideband modes. 
A direct measurement of both photocurrent Fourier quadrature components, if possible, would provide direct information on the \textit{optical} quadrature components of these specific modes~\cite{optcommEletQuad}.

\begin{figure}[ht]
\includegraphics[width=6cm]{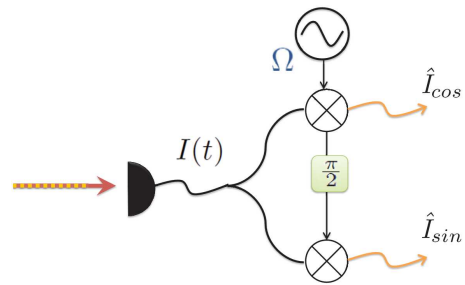}
\caption{Scheme to measure electronic quadrature components of each photocurrent signal. The photocurrent is mixed with two electronic references in quadrature. }
\label{fig:Icossin}
\end{figure}

In the ideal case, each measurement of an \textit{electronic} quadrature component thus represents a \textit{single-mode} measurement, free of assumptions.
We note that $[\hat I_{\cos{}},\hat I_{\sin{}}]=0$, as expected, since they represent independent observables. 
A possible  technique to perform this measurement is shown in Fig.~\ref{fig:Icossin}, by mixing the photocurrent with two electronic references in quadrature~\cite{shapirocossin}.

\subsection{Photocurrent measurement: Spectral noise power, stationarity, and the role of the phases}
\label{sec:spectralnoiseHD}

According to the  Wiener-Kintchine theorem,
for a stationary process, where the average of the two-time correlation function $\ave{I(t)I(t')}$ depends only on the difference of times $\tau=t'-t$, the spectral power $S(\Omega)$ is related to the correlation of signal Fourier components. In our case, relating the spectral power to the photocurrent operators results in 
\begin{equation}
\label{eq:Snoisespectrum}
%\int^\infty_{-\infty} 
\left\{
\begin{array}{c}
S(\Omega)=\ave{\hat I_\Omega\hat I_{-\Omega}}\,\qquad \mbox{and},\\
\ave{\hat I_\Omega\hat I_{\Omega'}}=0, \;\mbox{for}\; \Omega'\neq - \Omega
\end{array}
\right.
\end{equation}
where $\langle ... \rangle$ represents a quantum average.
 In particular, for a stationary process it imposes the condition 
%\begin{equation}
%\label{eq:Sstac}
$\ave{\hat I_\Omega\hat I_{\Omega}}=0$.
%\ave{\tilde I(\Omega)\tilde I(\Omega)}=0.
%\end{equation}
In what follows, when referring to a \textit{stationary quantum state}  we will mean \textit{a quantum state producing a 
photocurrent} satisfying Eq.~(\ref{eq:Snoisespectrum}).

The spectral noise power is proportional to the total energy present in the photocurrent quantum fluctuations. It retrieves a mixture of quadrature operator moments lacking phase information~\cite{cavesamplifiersPRA1982,ralphmixedPRA08}. When evaluated from the electronic quadrature components, it reads as
\begin{equation}
\label{eq:SIcosIsin}
S(\Omega)={\textstyle\frac{1}{2}}\ave{\hat I_\mathrm{cos}^2}+{\textstyle\frac{1}{2}}\ave{\hat I_\mathrm{sin}^2}.
\end{equation}
Thus the photocurrent noise power does not correspond to the second-order moment of a \textit{bona fide} mode operator in general~\cite{cavesamplifiersPRA1982}. However, it can be interpreted as a pure quadrature moment given certain assumptions about the quantum state, as discussed in Sec.~\ref{sec:phasemixingregime}.

On the other hand, in principle, we could gain more information about the quantum state by checking the stationarity of the photocurrent. 
In case it is stationary, it follows that 
\begin{equation}
\label{eq:stacIomega}
\ave{\hat I_\Omega \hat I_{\Omega}}=0 \Rightarrow 
\left\{
\begin{array}{c}
\Delta^2\hat I_\mathrm{cos}-\Delta^2\hat I_\mathrm{sin}=0,\\
\ave{\hat I_\mathrm{cos}\hat I_\mathrm{sin}}=0.
\end{array}
\right.
\end{equation}
Stationarity is equivalent to perfect \textit{symmetry} between the statistics of electronic quadrature components and \textit{lack of correlation} between them. We use this result many times throughout the paper.

The scenario above considers only the general procedure of using a bright LO to amplify the contribution of quantum modes of interest in the photocurrent quantum fluctuations. In order to achieve further insights, we must investigate the precise technique used to measure the field. 
The discussion presented here provides the fundamentals of the following analysis. We demonstrate next the incompleteness of homodyne detection, and how it can be overcome with the use of optical cavities for resonator detection.

\section{Spectral homodyne detection (HD)}
\label{sec:homodynedetection}

\begin{figure}[ht]
\includegraphics[width=4cm,angle=270]{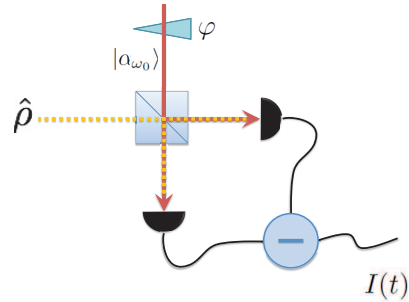}
\caption{Schematic view of the balanced homodyne detection. Prior to detection, LO field in state $\ket{\alpha_{\omega_0}}$ is added to the quantum field modes of interest with a controlled phase $\varphi$, using a 50/50 beam splitter. Information about the quantum field is retrieved after subtraction of the photocurrents. }
\label{fig:HD}
\end{figure}

Homodyne detection was the first technique to provide direct access to quadrature field observables, and still remains the most widely used measurement technique in the CV domain~\cite{shapirohd1,shapirohd2}. 
Balanced homodyne detection is the usual realization of HD in the laboratory~\cite{yuenchanOL1983,optlettschum,shapirorev1985} (see figure \ref{fig:HD}). The field modes to characterize are mixed on a 50/50 beamsplitter with the LO. Quantum measurement is obtained by the difference between photocurrents recorded on a pair of photodiodes placed on the two output ports of the beamsplitter. This scheme has the technical advantage of automatically canceling LO noise in detection. Non unity measurement efficiency can be taken into account by considering an ideal detector with a beam splitter in the path prior to detection~\cite{shapirocossin,jmo1987}. 

In order to measure the quantum state of spectral field modes (sideband modes), 
we perform the spectral analysis of the photocurrent quantum fluctuations. In this case, the technique is essentially described by the ideas presented in last section, and the treatment leading to the measurement operators of Eq.~(\ref{harmoniccomp}) can be directly applied. 
The spectral operator of spectral homodyne detection is 
\begin{equation}\label{Ibhd}
\hat I^{HD}_{\Omega}(\varphi) =e^{-i\varphi} \hat a_u+ e^{i\varphi}  \hat a^{\dagger}_l.
\end{equation}
Quantum state reconstruction follows from controlling the LO phase $\varphi$ to reveal different quadrature directions in the phase space of field modes. 
The phase can be mastered and easily varied at will~\cite{coherencefiction}. 

The electronic quadrature components of the spectral photocurrent represent each a pure single-mode measurement. 
The quadrature operators can be associated with the symmetric ($\mathcal{S}$) and anti-symmetric ($\mathcal{A}$) combinations of sidebands, as
\begin{equation}
%\left\{
\begin{array}{ccccl}
\hat I_\mathrm{cos}(\varphi)&=&\cos\varphi\,\hat p_s+\sin\varphi\,\hat q_s&\equiv& \hat X^\varphi_s,\\
\hat I_\mathrm{sin}(\varphi)&=&\cos\varphi\,\hat q_a-\sin\varphi\,\hat p_a&\equiv& \hat X^{\varphi+\frac{\pi}{2}}_a,
\end{array}
%\right.
\label{eq:HDcompcossinpm}
\end{equation}
where the quadrature observables
\begin{align}
\label{eq:defquadsymasym}
%\hat p_{s(a)}={\textstyle\frac{1}{\sqrt2}}(\hat p_u\pm\hat p_{\ell})\;\;\mathrm{and}\quad 
\hat p_{s(a)}=\frac{\hat p_u\pm\hat p_{\ell}}{\sqrt2}\quad\mathrm{and}\quad 
\hat q_{s(a)}=\frac{\hat q_u\pm\hat q_{\ell}}{\sqrt{2}}
\end{align}
represent the natural modes of the HD detection scheme, and $\hat X^\varphi_{s(a)}$ are corresponding generalized quadrature observables of these new modes.
In the modal basis of upper/lower sidebands, HD performs a Bell-type measurement.

Equation~(\ref{eq:HDcompcossinpm}) also shows that, although spectral HD is, in principle, able to provide two-mode operator moments, it can not achieve complete quantum state reconstruction. The reason for that is the fact that modes $\mathcal{S}$ and $\mathcal{A}$ can not be probed independently, since their measurement orientations
in phase space are fixed with respect to one another by a single parameter $\varphi$~\cite{prltobe}. 

\subsection{Spectral noise power and stationarity}
\label{subsec:stac}

As previously discussed, the spectral noise power consists in general of a mixture of modal operator moments~\cite{cavesamplifiersPRA1982,ralphmixedPRA08}. 
Using the measurement operators of Eq.~(\ref{eq:HDcompcossinpm}) in Eq.~(\ref{eq:SIcosIsin}), we find 
\begin{align}
\label{eq:SPmaismenos}
S_\mathrm{HD}(\varphi)&={\textstyle \frac{1}{2}}\Delta^2\hat X^\varphi_s+{\textstyle \frac{1}{2}}\Delta^2\hat X^{\varphi+\frac{\pi}{2}}_a.
\end{align}
In a more general perspective, the noise power 
is a direct experimental realization of 
 the Duan {\it et al.} entanglement criterion applied to sideband modes 
~\cite{simon,dgcz}, pointing at the well known fact that spectral quantum noise squeezing (i.e. $S_\mathrm{HD}<1$) witnesses two-mode sideband entanglement rather than a single-mode squeezed state of the field~\cite{huntington2002,zhangPRA2003,huntingtonPRA05}. 

Quantum noise corresponds to a pure quadrature moment only for a particular set of quantum states for which $\Delta^2\hat X^\varphi_s=\Delta^2\hat X^{\varphi+\frac{\pi}{2}}_a$. 
Whether the quantum state satisfies this property can, in principle, be independently verified by checking the \textit{stationarity of photocurrent fluctuations} in the experiment [Eq.~(\ref{eq:stacIomega})]. In HD, this condition implies 
\begin{equation}
\label{eq:stacIomegaHD}
\ave{\hat I_\Omega^{HD} \hat I_{\Omega}^{HD}}=0 \Rightarrow 
\left\{
\begin{array}{c}
\Delta^2\hat X^\varphi_s=\Delta^2\hat X^{\varphi+\frac{\pi}{2}}_a\\
\ave{\hat X^\varphi_s\hat X^{\varphi+\frac{\pi}{2}}_a}=0
\end{array}
\right..
\end{equation}
We are then led to the result that \textit{in the special case of a stationary quantum state} [Eq.~(\ref{eq:stacIomega})], \textit{and only in this case, the noise power indeed corresponds to the variance of a proper field mode quadrature}. In this case, it can be interpreted either as a single-mode measurement of the symmetric ($\mathcal{S}$) or the anti-symmetric ($\mathcal{A}$) combination of sideband modes, since their quantum states are essentially the same, differing only by a local rotation, as seen below. 

\subsection{Covariance matrix for stationary quantum states}
\label{subseccovstac}

Observation of stationary photocurrent in spectral HD reveals certain aspects of the quantum state, imposing constraints on the covariance matrix. To satisfy stationarity [Eq.~(\ref{eq:stacIomegaHD})], the covariance matrix written in the $\mathcal{S}/\mathcal{A}$ modal basis must assume the highly symmetric form~\cite{cavesamplifiersPRA1982} 
\begin{equation}
\mathbf{V}_{(s/a)}=
\left(
\begin{array}{cccc}
\alpha & \gamma & \delta & 0\\
\gamma & \beta & 0 & \delta\\
\delta & 0 & \beta & -\gamma\\
0 & \delta & -\gamma & \alpha
\end{array}
\right)
\equiv\left(
\begin{array}{cc}
\mathbf{V}_s & \mathbf{C}_{(s/a)} \\
(\mathbf{C}_{(s/a)})^T & \mathbf{V}_a \\
\end{array}
\right),
\label{eq:Vstat}
\end{equation}
where $\mathbf{V}_{(s/a)}=\ave{[\vec X_{s},\vec X_{a}]\cdot[\vec X_{s},\vec X_{a}]^T}$,  with quadrature operators arranged in a vector as $\vec X_{s(a)}=(\hat p_{s(a)} \; \hat q_{s(a)})^T$. We have defined the single-mode covariance matrices of $\mathcal{S}$ and $\mathcal{A}$ modes as $\mathbf{V}_{s(a)}=\ave{\vec X_{s(a)}\cdot\vec X_{s(a)}^T}$. The matrix $\mathbf{C}_{(s/a)}=\ave{\vec X_s\cdot\vec X_a^T}$ contains two-mode correlations.

Stationarity hence implies that modes $\mathcal{S}$ and $\mathcal{A}$ present equal quantum statistics (or, equivalently, possess  the same local quantum state) apart from a local rotation of quadratures, i.e. $\Delta^2\hat p_s=\Delta^2\hat q_a\equiv\alpha$, $\Delta^2\hat q_s=\Delta^2\hat p_a\equiv\beta$ for the variances and $C(\hat p_s\hat q_s)=-C(\hat p_a\hat q_a)\equiv\gamma$ for the correlations. In other words, $\mathbf{V}_s$ is equal to $\mathbf{V}_a$ after a rotation of $\pi/2$ on the quadrature phase space of one of the modes. 
Nevertheless, two-mode correlations can still be present in stationary states, through the correlation moment $C(\hat p_s\hat p_a)=C(\hat q_s\hat q_a)\equiv\delta$.

Consequences of stationarity can also be analyzed in the modal basis of lower and upper sidebands, in which case the covariance matrix is obtained from Eq.~(\ref{eq:Vstat}) by a simple rotation of quadratures [Eq.~(\ref{eq:defquadsymasym})]. It assumes the general symmetric form
\begin{equation}
\mathbf{V}_{(\ell/u)}=
\left(
\begin{array}{cccc}
\alpha' & 0 & \gamma' & \delta'\\
0 & \alpha' & \delta' & -\gamma'\\
\gamma' & \delta' & \beta' & 0\\
\delta' & -\gamma' & 0 & \beta'
\end{array}
\right)
\equiv
\left(
\begin{array}{cc}
\mathbf{V}_{\ell} & \mathbf{C}_{(\ell/u)} \\
(\mathbf{C}_{(\ell/u)})^T & \mathbf{V}_{u} \\
\end{array}
\right),
\label{eq:VstationarySB}
\end{equation}
where $\mathbf{V}_{\ell}$ and $\mathbf{V}_{u}$ are the covariance matrices of individual sideband modes, defined as $\mathbf{V}_{u}=\ave{\vec X_{u}\cdot\vec X_{u}^T}$, with $\vec X_{u}=(\hat p_{u} \; \hat q_{u})^T$ (analogously for mode $\ell$), and $\mathbf{C}_{(\ell/u)}=\ave{\vec X_{\ell}\cdot\vec X_{u}^T}$ contains sideband correlations. 

In the basis of sideband modes, quantum state symmetry manifests itself by the identities $\Delta^2\hat p_{\ell}=\Delta^2\hat q_{\ell}=\alpha'$, $\Delta^2\hat p_{u}=\Delta^2\hat q_{u}=\beta'$, $C(\hat p_{\ell}\hat p_{u})=-C(\hat q_{\ell}\hat q_{u})=\gamma'$ and $C(\hat p_{\ell}\hat q_{u})=C(\hat q_{\ell}\hat p_{u})=\delta'$. 

Thus, stationarity implies that sideband modes are in thermal states, but may show correlations, leading to entanglement depending on the amount of shared information. In the specific case of a two-mode \textit{pure state}, sideband modes producing a stationary photocurrent are either in the vacuum state or form an entangled EPR-like state.

\subsection{Incomplete quantum state reconstruction of stationary quantum states}

We show now that the two  pieces of information ideally available experimentally, namely the spectral noise power and the stationarity of the photocurrent, are not sufficient to determine the most general spectral two-mode quantum state in HD. 

Indeed, using Eq.~(\ref{eq:HDcompcossinpm}), the spectral noise power of HD [Eq.~(\ref{eq:SPmaismenos})] combines the moments of two modes as
\begin{align}
\label{eq:imixexplicit}
%S_\mathrm{HD}&=  \frac{\cos^2\varphi}{2}\left(\Delta^2\hat p_++\Delta^2\hat q_-\right)+\frac{\sin^2\varphi}{2}\left(\Delta^2\hat p_-+\Delta^2\hat q_+\right)\nonumber\\
%&+\frac{\sin2\varphi}{2}\left(C_{\hat p_+\hat q_+}-C_{\hat p_-\hat q_-}\right).
S_\mathrm{HD}(\varphi)=&  \cos^2\varphi\;\frac{\Delta^2\hat p_s+\Delta^2\hat q_a}{2} +\sin^2\varphi\;\frac{\Delta^2\hat p_a+\Delta^2\hat q_s}{2}\nonumber\\
&+\sin2\varphi\;\frac{C(\hat p_s\hat q_s)-C(\hat p_a\hat q_a)}{2}.
\end{align}
Owing to stationarity [Eq.~(\ref{eq:stacIomegaHD})], the noise power simplifies to a single-mode expression. Written in terms of the elements of the covariance matrix [Eq.~(\ref{eq:Vstat})], it reads as
\begin{align}
S_\mathrm{HD}(\varphi)=\cos^2\!\varphi\;\alpha+\sin^2\!\varphi\;\beta+\sin2\varphi\;\gamma,
\label{eq:SHDstac}
\end{align}
whereby it becomes clear that the moment $\delta=C(\hat p_s\hat p_a)=C(\hat q_s\hat q_a)$ of a \textit{general stationary quantum state} is \textit{missing}. 
The physical significance of the missing moment is better realized in the modal basis of sidebands, where 
$2\delta=\alpha'-\beta'=%{\displaystyle\frac{1}{2}}
(\Delta^2\hat p_{u}+\Delta^2\hat q_{u})-(\Delta^2\hat p_{\ell}+\Delta^2\hat q_{\ell})$: \textit{It yields the energy imbalance between sideband modes}~\cite{prltobe}.  

The intrinsic insensitivity of HD to modal energy imbalance should be expected from the symmetry with which it treats sideband modes, making it impossible to detect sideband asymmetry. Upper and lower sidebands are completely indistinguishable from one another in the spectral noise power of HD, as seen in Eq.~(\ref{eq:HDcompcossinpm}). 
The same equation on the $\mathcal{S}/\mathcal{A}$ modal basis shows that $\hat X^{\varphi}_a$ and $\hat X^{\varphi'}_s$ can not be measured independently of one another, since HD imposes $\varphi'=\varphi+\pi/2$. This fact hinders the complete reconstruction of $\mathcal{S}/\mathcal{A}$ two-mode correlation, represented by $\delta$.

Since $\delta$ is inaccessible by HD, it must be implicitly assumed as null in most quantum state reconstruction experiments ($\delta=0\Rightarrow\alpha'=\beta'$). This assumption of {\it a priori} knowledge about the quantum state is in many cases reasonable, e. g. in squeezed state generation by spontaneous parametric downconversion (SPDC)~\cite{squeezing}, due to the broadband nature of emission. Nevertheless, in more complex systems, this term could be important to reveal entanglement among sidebands~\cite{prltobe}. In particular, resonant phenomena such as atomic emission should lead to strong energy asymmetry among longitudinal modes. 

Hence spectral HD applied to a single beam is an intrinsically single-mode measurement technique, limited to the reconstruction of an {\it effective single-mode }
for stationary quantum states. 
This mode can be either regarded as the symmetric or anti-symmetric combination of sideband modes, since they bear the same quantum state in the case of stationary photocurrent signals. 
For this measurement to be complete, in addition to the stationarity condition, one has to assume the $\mathcal{S}/\mathcal{A}$ modes to be uncorrelated or, equivalently, that sideband modes carry the same mean energy. We now see that a complete measurement, free from such a limiting assumption, is possible with the resonator detection technique.

\section{Resonator detection (RD)}
\label{sec:resonatordetection}

We now examine the measurement operator associated with resonator detection~\cite{levenson,galatola,zhangJOSAB2000,zavattaPRA2002,villarajp}. 
The technique is based on the dispersive property of an optical resonator close to resonance, bringing an intrinsic asymmetry in the way sideband modes are manipulated before photodetection. 
It has been employed to measure quantum noise squeezing in the pioneering work by Shelby \textit{et al.}~\cite{shelby}.

\begin{figure}[ht]
\includegraphics[width=4cm,angle=270]{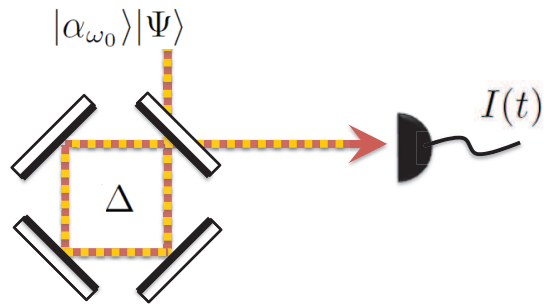}%{RD.eps}
\caption{Schematic view of resonator detection. The state of interest and the carrier mode  are reflected off an optical resonator prior to photodetection. Frequency dependent losses and phase shifts, controlled by resonator detuning $\Delta$, allow quantum state reconstruction. }
\label{fig:RD}
\end{figure}

Resonator detection is realized by measuring the intensity fluctuation of a field after its reflection off an optical resonator, as schematized in Fig.~\ref{fig:RD}. Field modes in a narrow band close to resonance with the optical cavity are phase-shifted and attenuated just prior to detection. Similarly to HD, RD needs an intense LO field to amplify sideband mode quantum fluctuations in detection. A convenient displacement operator can be applied prior to the cavity operation if the state to be measured is `dim'.

An optical resonator with high finesse transforms the field annihilation operators according to~\cite{galatola,villarajp}
\begin{equation}
\hat a_{\omega}\longrightarrow r(\Delta_\omega)\,\hat a_{\omega} + t(\Delta_\omega)\,\hat b_{\omega},
\label{eq:aomegaref}
\end{equation}
where $r(\Delta_\omega)$ and $t(\Delta_\omega)=\sqrt{1-r^2(\Delta_\omega)}$ are respectively resonator reflection and transmission frequency responses. A vacuum mode in transmission, described by the annihilation operator $\hat b_{\omega}$, substitutes the missing fraction of reflected modes, a feature which proves essential to the power of the technique regarding quantum state reconstruction.

Reflection induces frequency-dependent phase shift and loss, as functions of the detuning $\Delta_\omega=(\omega-\omega_c)/\gamma$ between longitudinal mode frequency $\omega$ and resonator frequency $\omega_c$ ($\gamma$ is the resonator bandwidth).
Close to one given resonance, its explicit form is 
\begin{equation}
r(\Delta_\omega)=-\frac{\sqrt{d}+2i\Delta_\omega}{1-2i\Delta_\omega},
\label{eq:defrDelta}
\end{equation} 
where $d$, the impedance matching parameter, is the fraction of reflected light at exact resonance ($d = |r(0)|^2$). 
It depends on the ratio between input mirror coupling and resonator losses. In the extreme cases, an ideal lossless resonator has $d=1$ (input beam is totally reflected), while $d=0$ indicates a so-called `impedance matched resonator' (the spectral mode reflected at exact resonance is completely substituted by a transmitted mode in vacuum state). 
The LO mode in particular undergoes the transformation
\begin{equation}
\alpha\longrightarrow r(\Delta)\,\alpha,
\label{eq:aomegaLO}
\end{equation}
where $\Delta=(\omega_0-\omega_c)/\gamma$ is the detuning $\Delta$ between carrier and resonator frequency. We consider the initial carrier phase to be zero, i.e. $\alpha=\alpha^*$ without loss of generality. The detuning $\Delta$ is the experimentally controllable parameter of RD. %(MM)

In RD, the general photocurrent operator [Eq.~(\ref{eq:Iomeganonhermitian})] is modified to include not only a dephasing of LO mode [Eq.~(\ref{eq:aomegaLO})], as in HD, but also a unitary transformation acting on the annihilation/creation operators of quantum modes nearly resonant with the optical cavity [Eq.~(\ref{eq:aomegaref})]. 
Substituting the operators and carrier amplitude of Eqs.~(\ref{eq:aomegaref}) and~(\ref{eq:aomegaLO}) in Eq.~(\ref{eq:Iomeganonhermitian}), the spectral operator of RD reads as
\begin{eqnarray}
\label{eq:Jomegaloss}
\hat J_\Omega(\Delta)&=&R_\Omega^*(\Delta)\;\hat a_u+R_{-\Omega}(\Delta)\;\hat a^\dag_{\ell} \nonumber \\
&\, & +T_\Omega^*(\Delta)\;\hat b_u+T_{-\Omega}(\Delta)\;\hat b^\dag_{\ell}, %\nonumber \\
\end{eqnarray}
where the $\Delta$-dependent coefficients are 
\begin{align}
R_{\Omega}(\Delta)&=\frac{1}{\sqrt2}\frac{r(\Delta)}{|r(\Delta)|}\;r^*(\Delta+\Omega/\gamma),\nonumber\\ 
T_{\Omega}(\Delta)&=\frac{1}{\sqrt2}\frac{r(\Delta)}{|r(\Delta)|}\;t^*(\Delta+\Omega/\gamma). 
\end{align}

The operator of Eq.~(\ref{eq:Jomegaloss}) represents in a concise notation the two Hermitian measurement operators for the electronic quadrature components of the photocurrent $\hat J_\mathrm{cos}$ and $\hat J_\mathrm{sin}$, together with vacuum terms due to depletion the sidebands undergo when resonant ($\hat J_{vac}$), by the expression $\hat J_\Omega = (\hat J_\mathrm{cos} + i \hat J_\mathrm{sin})/\sqrt2 + \hat J_{vac}$. Disregarding vacuum terms for the moment, the Hermitian measurement operators are 
\begin{align}
\label{eq:Jcossin}
\left\{
\begin{array}{ccc}
\hat J_\mathrm{cos}(\Delta) &=&x_+\hat p_u+y_+\hat q_u+x_-\hat p_\ell -y_-\hat q_\ell \\
\hat J_\mathrm{sin}(\Delta) &=&y_-\hat p_u+x_-\hat q_u-y_+\hat p_\ell +x_+\hat q_\ell
\end{array}
\right.,
\end{align}
%\begin{align}
%\label{eq:Jcos}
%\hat J_\mathrm{cos}(\Delta) &=x_+\hat p_++y_+\hat q_++x_-\hat p_--y_-\hat q_- ,\\
%\label{eq:Jsin}
%\hat J_\mathrm{sin}(\Delta) &=y_-\hat p_++x_-\hat q_+-y_+\hat p_-+x_+\hat q_-
%\end{align}
where $x_\pm$ and $y_\pm$ are real functions of $\Delta$ defined as
\begin{align}
x_++iy_+&=(R_{\Omega}+R^*_{-\Omega})/2\equiv g_+ \,,\nonumber \\
x_-+iy_-&=i(R_{\Omega}-R^*_{-\Omega})/2\equiv g_- \, .
\label{defineg}
\end{align}
We note that $[\hat J_\mathrm{cos},\hat J_\mathrm{sin}]=0$, since they represent independent quantum observables.

Contrarily to the case of HD, the electronic quadrature measurement operators $\hat J_\mathrm{cos}$ and $\hat J_\mathrm{sin}$ of RD undergo changes of modal basis depending on $\Delta$, revealing the inherent two-mode character of the technique.

\subsection{Spectral noise power and complete state reconstruction of stationary quantum states}

The photocurrent spectral noise power of RD is obtained from Eqs.~(\ref{eq:SIcosIsin}) and~(\ref{eq:Jcossin}), yielding
\begin{equation}
S_\mathrm{RD}=\ave{\hat J_\Omega\hat J_{-\Omega}}={\textstyle\frac{1}{2}}\Delta^2\hat J_\mathrm{cos}+{\textstyle\frac{1}{2}}\Delta^2\hat J_\mathrm{sin}+ \Delta^2 \hat J_\mathrm{vac},
\label{eq:noiseJ}
\end{equation}
where $\Delta^2 \hat J_\mathrm{vac}$ is the vacuum noise contribution.

\begin{figure}[ht]
\includegraphics[width=0.95\linewidth]{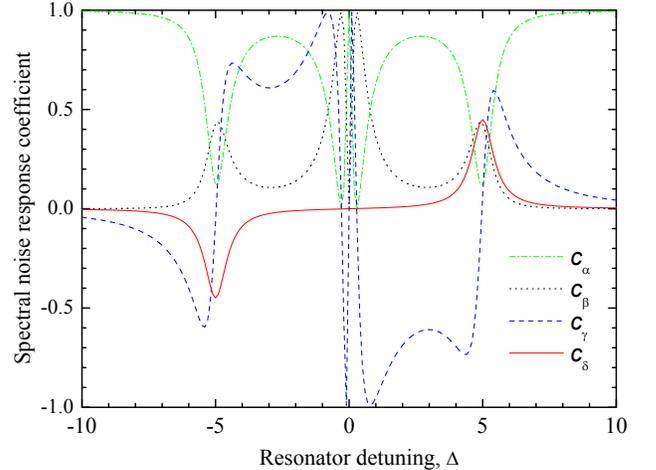}
\caption{(Color online) Coefficients of Eq.~(\ref{eq:Srd}) as functions of $\Delta$. Resonator parameters are $d=0.9$ and $\Omega/\gamma=5$.}
\label{fig:coefsrotelipse}
\end{figure}

Stationarity of electronic quadrature components in RD results in the same considerations of Sec.~\ref{subsec:stac} and hence imposes for the covariance matrix the forms of Eqs.~(\ref{eq:Vstat})--(\ref{eq:VstationarySB}). Explicitly writing Eq.~(\ref{eq:noiseJ}) in terms of moments of quadrature observables with the aid of Eq.~(\ref{eq:Jcossin}), we find the spectral quantum noise of resonator detection in terms of elements of the covariance matrix [Eq.~(\ref{eq:Vstat})] as 
\begin{equation}
S_\mathrm{RD}(\Delta)=c_\alpha\,\alpha +c_\beta\,\beta+c_\gamma\,\gamma +c_\delta\delta+ c_v\, ,
%S_\mathrm{RD}(\Delta)=c_\alpha(\Delta)\,\alpha +c_\beta(\Delta)\,\beta+c_\gamma(\Delta)\,\gamma +c_\delta(\Delta)\delta+ c_v(\Delta)\, ,
\label{eq:Srd}
\end{equation}
where $c_\alpha=|g_+|^2$, $c_\beta=|g_-|^2$, $c_\gamma+ic_\delta=2g^*_+g_-$ and $c_v=1-c_\alpha-c_\beta$ are functions of $\Delta$.
This expression shows that the spectral noise power of RD reveals \textit{all four second-order moments in the covariance matrix} described in Eq.~(\ref{eq:Vstat}) needed to determine any stationary two-mode Gaussian quantum state. 
Figure~\ref{fig:coefsrotelipse} confirms that each term of Eq.~(\ref{eq:Srd}) has a distinct dependence on resonator detuning $\Delta$, allowing one to distinguish the  contribution of each individual quadrature moment in the spectral quantum noise~\cite{prltobe}.

The sensitivity of $S_\mathrm{RD}$ to each operator moment depends on two experimental parameters: First, the analysis frequency $\Omega$, which must be larger than $\sqrt{2}\gamma$ to allow better access to phase quadrature moments~\cite{villarajp}; second, the impedance matching parameter $d$ [Eq.~(\ref{eq:defrDelta})]. 

In particular, the sensitivity to the `hidden moment' $\delta$, as determined by the coefficient $c_\delta$, is maximum for $d=0$.  
At exact resonance with one sideband, the impedance matched resonator maximizes the response asymmetry to its longitudinal counterpart. By substituting one sideband by a field mode in vacuum state, the resonator separates sidebands spatially~\cite{huntingtonPRA05}. 

On the other extreme, an ideal lossless resonator ($d=1$) acts as a simple phase shifter, simply dephasing LO with respect to sidebands as in HD. In this limit, RD provides a HD-like measurement and hence becomes insensitive to the `hidden moment' $\delta$. It is then possible to write $c_\alpha \equiv\cos^2\varphi$, $c_\beta\equiv\sin^2\varphi$, $c_\delta=0$ and $c_\gamma=2\sin\varphi$, where the phase shift $\varphi$ is a function of detuning $\Delta$~\cite{villarajp}. In other words, Eq.~(\ref{eq:Srd}) reduces exactly to Eq.~(\ref{eq:SHDstac}), showing that the essential feature that distinguishes HD and RD is the way sideband modes contribute to quantum noise. In RD, resonator detuning varies not only the phase of spectral modes with respect to LO, but also the relative amount of modal contribution to quantum noise. Hence the absolute values of coefficients in Eq.~(\ref{eq:Srd}) play a crucial role in achieving complete state reconstruction.

\section{Phase mixing regime}
\label{sec:phasemixingregime}

Until this point, we have treated the electronic photocurrent sine and cosine components as the measurement operators associated with HD and RD detection schemes. We have established the distinction between these techniques regarding their capacity to reconstruct stationary Gaussian quantum states, in particular pointing to the limitations of HD to determine some classes of two-mode quantum states. 

Although these photocurrent components are, in principle, retrievable by measurement, they require a common phase reference between the \textit{optical} LO and the \textit{electronic local oscillator} (eLO) used to extract the desired photocurrent Fourier component~\cite{jmo1987,naturebreitenbach}. However, in a typical experimental situation, the optical LO shows relatively fast phase diffusion~\cite{phaselockingPRA10}.
If the laser linewidth is not narrow enough to allow a complete characterization of the state before phase diffusion becomes important, or if it is not phase locked to the electronic oscillator, the measurement operator will vary between individual quantum measurements, introducing mixedness in the photocurrent moments. 

We can analyze this case with a simple model.  If we consider a linear combination of cosine and sine electronic quadrature components, in the form
\begin{equation}
\hat I_\theta=\cos\theta\,\hat I_\mathrm{cos}+\sin\theta\,\hat I_\mathrm{sin} \;,
\label{eq:imix}
\end{equation}
we may conceive that the relative phase $\theta$ between LO and eLO remains constant during a single quantum measurement but varies during the collection of quantum statistics. 

In this case, moments of photocurrent fluctuations are obtained by $\theta$-averages  of moments of $\hat I_\theta$. 
Regarding second order moments, the variance of any \textit{measured} photocurrent component becomes a mixture of variances of cosine and sine components, since 
\begin{align}
\label{eq:varsincosnull}
\Delta^2 \hat I_\theta&={\textstyle\frac{1}{2\pi}}\int d\theta'\,\ave{\hat I_{\theta+\theta'}\hat I_{\theta+\theta'}}\nonumber\\
&={\textstyle \frac{1}{2}}\Delta^2 \hat I_{\cos{}}+{\textstyle \frac{1}{2}}\Delta^2\hat I_\mathrm{sin}, \;\forall\theta.
\end{align}
Furthermore, correlation between in quadrature photocurrent components of a single beam is always zero, since
\begin{align}
C_{\hat I_\theta\hat I_{\theta+\pi/2}}& ={\textstyle\frac{1}{2\pi}}\int d\theta'\ave{\hat I_{\theta+\theta'}\,\hat I_{\theta+\frac{\pi}{2}+\theta'}}\nonumber\\
&=\ave{\hat I_\mathrm{cos}\hat I_\mathrm{sin}-\hat I_\mathrm{sin}\hat I_\mathrm{cos}}=0, \forall\theta,
%{\textstyle \frac{1}{2}}\int d\theta \cos\theta\sin\theta\ave{\hat I_\mathrm{cos}\hat I_\mathrm{sin}-\hat I_\mathrm{sin}\hat I_{\cos{}}}=0.
\label{eq:corrsincosnull}
\end{align}
where we have used $[\hat I_{\cos{}},\hat I_\mathrm{sin}]=0$, independently of the quantum state.

The conditions above, implied by phase mixing and valid for both HD and RD (substituting $\hat I_\theta\leftrightarrow\hat J_\theta$), are summarized as
\begin{align}
\label{eq:Iphasemix}
\left\{
\begin{array}{ccl}
\Delta^2 \hat I_\theta&=&\Delta^2 \hat I_{\theta+\frac{\pi}{2}},\\
C_{\hat I_\theta\hat I_{\theta+\frac{\pi}{2}}}&=&0.
\end{array}
\right.
\end{align}

Hence in the context of phase mixing,  any measured $\theta$ photocurrent component should present the same statistics and be uncorrelated, i.e. all information available must lie in any single and arbitrary photocurrent component. 

As a matter of fact, the properties imposed by Eq.~(\ref{eq:Iphasemix}) on the measured photocurrent coincide with the conditions for stationarity, according to Eq.~(\ref{eq:stacIomega}).
Thus in the phase mixing scenario, the photocurrent is always stationary regardless of the quantum state of light. 
In this scenario, the conditions of Eq.~(\ref{eq:stacIomega}) cannot be applied to infer elements of the covariance matrix and bring it to the form of Eq.~(\ref{eq:Vstat}), since stationarity could be just a consequence of phase mixing, and not a property of the quantum state. The spectral noise power then stands as the only experimentally meaningful signal available. 

In order to obtain more information about the field modes, one needs to recover properties of the quantum state \textit{subjacent to phase mixing}. For instance by determining whether the sine and cosine electronic quadrature components are stationary themselves. It turns out that higher order moments of the measured photocurrent yield the desired information in the case of Gaussian states~\cite{gaussiantobe}. In our measurements, we are thus able to establish stationarity of any Gaussian quantum state by indirect means despite phase mixing. In the following, we treat only stationary quantum states.

\section{Generalization to more beams}
\label{sec:generalization}

We now consider the problem of determining the Gaussian quantum state of any number of beams. 
Joint measurements are necessary to reconstruct the collective multimode quantum state in this case. For Gaussian states, second-order moments suffice to describe the global system, so that only pairwise correlations determine the multimode state. Hence the collective quantum state of any number of beams is determined by reconstructing the state of every possible pair independently. 

We treat here the fundamental building block of multimode measurement, by explicitly providing the generalization of previous sections to two beams, i.e. four optical modes distributed as two longitudinal sideband modes per beam. For the sake of concreteness, and to facilitate the comparison with measured quantities in Section~\ref{sec:experimental}, we consider in this section the photocurrent moments as measured for stationary quantum states. 

Given the primacy of symmetric ($\mathcal{S}$) and anti-symmetric ($\mathcal{A}$) modes in the measurement of quantum noise of a single beam [Eq.~\ref{eq:SPmaismenos}], we write here the four-mode covariance matrix for two beams in this modal basis as $\mathbf{V}_{(s/a)}^{(12)}=\ave{(\vec X_{s}^{(12)}, \vec X_{a}^{(12)})\cdot (\vec X_{s}^{(12)}, \vec X_{a}^{(12)})^T}$, 
where the vector $\vec X_{s}^{(12)}= (\hat p_{s}^{(1)}\;\hat q_{s}^{(1)}\;\hat p_{s}^{(2)}\;\hat q_{s}^{(2)})$ involves the symmetric combination of sideband modes respective to modes of beams $(1)$ and $(2)$. Vectors for the anti-symmetric modes are defined analogously, as $\vec X_{a}^{(12)}= (\hat p_{a}^{(1)}\;\hat q_{a}^{(1)}\;\hat p_{a}^{(2)}\;\hat q_{a}^{(2)})$.
With this arrangement, the covariance matrix assumes the form
\begin{equation}
\mathbf{V}_{(s/a)}^{(12)}=\left(
\begin{array}{cc}
\mathbf{V}_{s}^{(12)} & \mathbf{C}_{(s/a)}^{(12)} \\
(\mathbf{C}_{(s/a)}^{(12)}) ^T & \mathbf{V}_{a}^{(12)}
\end{array}
\right),
\label{eq:V4mode}
\end{equation}
where $\mathbf{V}_{s}^{(12)}$ ($\mathbf{V}_{a}^{(12)}$) collects only symmetric (anti-symmetric) moments of beams $(1)$ and $(2)$,
and $\mathbf{C}_{(s/a)}^{(12)}$ refers to correlations among $\mathcal{S}$ modes on one beam and $\mathcal{A}$ modes on the other. 

For stationary quantum states, as shown in Sec.~\ref{subseccovstac}, the two-beam covariance matrix $\mathbf{V}_{s}^{(12)}$ of symmetric modes is equal to its antisymmetric counterpart $\mathbf{V}_{a}^{(12)}$ up to a local basis rotation, and it assumes the explicit form
\begin{align}
\mathbf{V}_{s}^{(12)}=
\left(
\begin{array}{cccc}
\alpha^{(1)} & \gamma^{(1)} & \mu & \xi\\
\gamma^{(1)} & \beta^{(1)} & \zeta & \nu\\
\mu & \zeta & \alpha^{(2)} & \gamma^{(2)}\\
\xi & \nu & \gamma^{(2)} & \beta^{(2)}
\end{array}
\right)
%\nonumber\\ &
=\left(
\begin{array}{cc}
\mathbf{V}_{s}^{(1)} & \mathbf{C}_s^{(12)} \\
\mathbf{C}_s^{(21)} & \mathbf{V}_{s}^{(2)}
\end{array}
\right)
\label{eq:Vplus12Vminus12}
\end{align}

The covariance matrix $\mathbf{V}_{s}^{(12)}$ is composed of three distinct $2\times2$ blocks. The diagonal blocks are covariance matrices of individual modes $s^{(j)}$, identified in Eq.~(\ref{eq:Vstat}), and the off-diagonal block stands for the cross correlations between the symmetric modes of the beams. The two-mode covariance matrix $\mathbf{V}_{a}^{(12)}$ of anti-symmetric modes has a similar structure, up to a local phase rotation.

In order to determine the complete four-mode quantum state, we are left to consider the correlation matrix $\mathbf{C}_{(s/a)}^{(12)}$ [off-diagonal matrix of Eq.~(\ref{eq:V4mode})], which assumes for stationary quantum states the explicit form 
\begin{equation}
\mathbf{C}_{(s/a)}^{(12)}=\left(
\begin{array}{cccc}
\delta^{(1)} & 0  & \kappa & -\eta \\
0 & \delta^{(1)} & \tau & -\lambda \\
-\lambda & \eta & \delta^{(2)} & 0 \\
-\tau & \kappa & 0 & \delta^{(2)}
\end{array}
\right).
\label{eq:C4modest}
\end{equation}
Two $2\times2$ blocks in the diagonal consist of single-beam operator moments that are `hidden' to HD, as seen previously, consisting of same beam $\mathcal{S}^{(j)}/\mathcal{A}^{(j)}$ correlation. The off-diagonal matrix refers to correlations between $\mathcal{S}^{(j)}/\mathcal{A}^{(j')}$ modes ($j\neq j'$). 
Conversion to the sideband modal basis is straightforward, obtained in the same manner as changing modal basis between Eqs.~(\ref{eq:Vstat}) and~(\ref{eq:VstationarySB}), by employing the modal basis transformation of Eq.~(\ref{eq:defquadsymasym}).

Finally, we point out that to extend quantum state reconstruction to all longitudinal modes, one would have to scan LO frequency to cover a bandwidth of interest and record the quantum noise over a wide range of analysis frequencies. The resulting data would give complete information about the longitudinal multimode quantum state of a single beam, enabling the reconstruction of the two-time correlation matrix $V(\tau)$ via Fourier transform.

\subsection{Two-beam photocurrent correlations in the phase mixing regime}

To reconstruct the complete four-mode stationary quantum state of two beams one needs to determine, in addition to the two-mode longitudinal covariance matrix of individual beams, the eight two-beam correlation moments of Eqs.~(\ref{eq:Vplus12Vminus12}) and~(\ref{eq:C4modest}). 

\begin{figure}[ht]
\includegraphics[width=6cm]{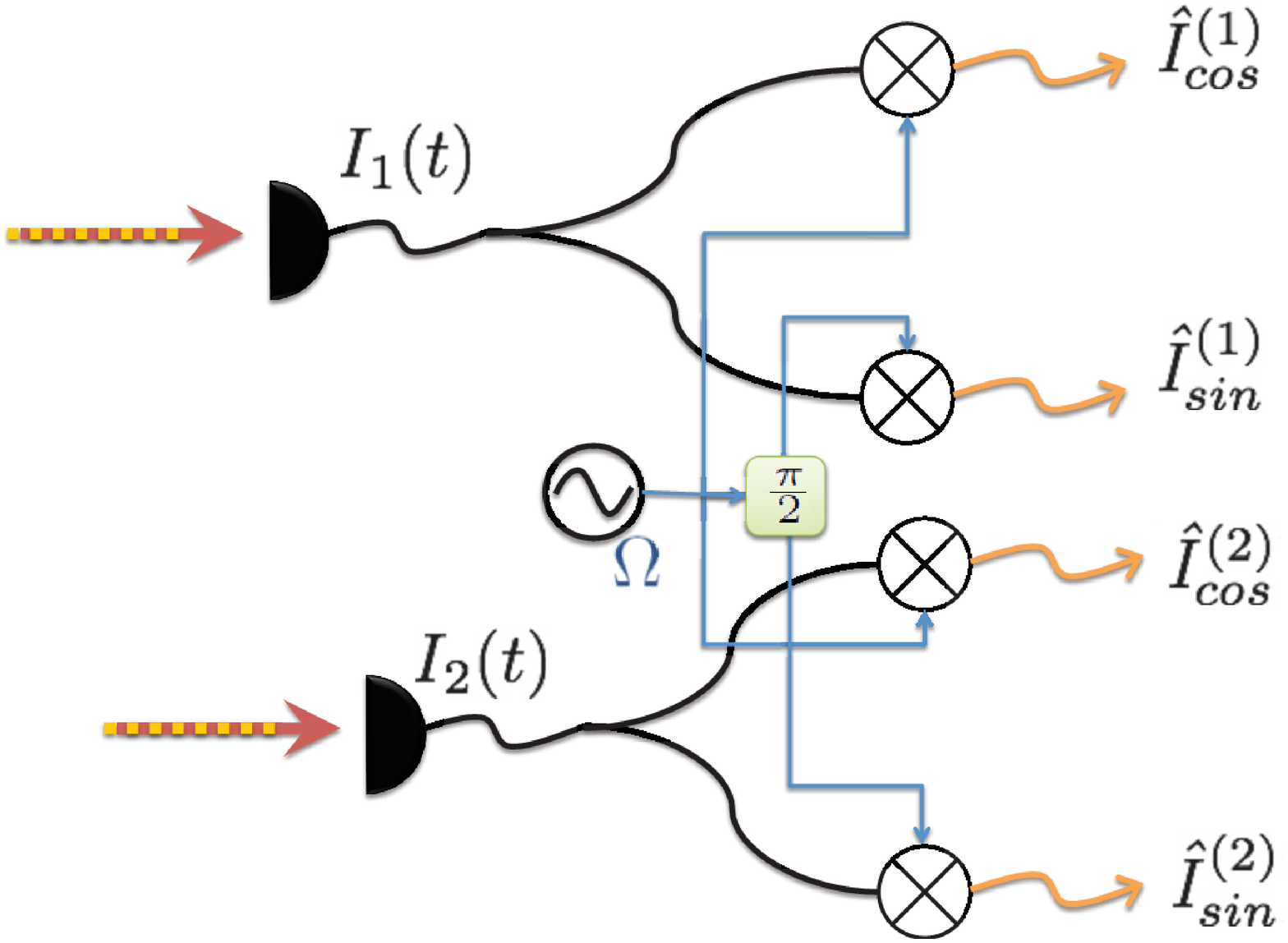}
\caption{Scheme to measure electronic quadrature components of two photocurrent signals produced by a pair of light beams. Photocurrents are mixed with two electronic references in quadrature. }
\label{fig:Icossin2}
\end{figure}

To achieve that, four experimental signals are available in the measurement of two beams, consisting of two photocurrent components for each beam (Fig.~\ref{fig:Icossin2}). 

We denote electronic quadrature photocurrent components of each beam by the measurement operators $\hat I_\mathrm{cos}^{(j)}$ and $\hat I_\mathrm{sin}^{(j)}$ in HD and $\hat J_\mathrm{cos}^{(j)}$ and $\hat J_\mathrm{sin}^{(j)}$ in RD. Variances of these operators result in individual noise spectra for both beams, denoted as $S_\mathrm{HD}^{(j)}$ [Eq.~(\ref{eq:SHDstac})] and $S_\mathrm{RD}^{(j)}$ [Eq.~(\ref{eq:Srd})]. They provide information about the covariance matrices of $\mathbf{V}_{s}^{(j)}$ and $\mathbf{V}_{a}{}^{(j)}$
%=\mathrm{Re}\{\mathbf{S}^{(j)}\}$ [Eq.~(\ref{eq:SVplusVminus})] 
in each technique. Sideband energy imbalance $\delta^{(j)}$ of each beam can be measured only with RD.
The remaining moments, involving cross correlations of electronic quadratures issued from photocurrents generated by different beams, are now examined. Stationarity, in this case, assures that
\begin{align}
\ave{\hat I_\mathrm{cos}^{(1)}\hat I_\mathrm{cos}^{(2)}}&= \ave{\hat I_\mathrm{sin}^{(1)}\hat I_\mathrm{sin}^{(2)}}, \\
\ave{\hat I_\mathrm{sin}^{(1)}\hat I_\mathrm{cos}^{(2)}} &= -\ave{\hat I_\mathrm{cos}^{(1)}\hat I_\mathrm{sin}^{(2)}}.
\end{align}
These terms are related to the cross correlation of spectral photocurrent components by 
\begin{align}
\label{eq:Cphasedef}
\mathrm{Re}\{\ave{\hat I_\Omega^{(1)}\hat I_{-\Omega}^{(2)}}\}&={\textstyle\frac{1}{2}}
\ave{\hat I_\mathrm{cos}^{(1)}\hat I_\mathrm{cos}^{(2)}}+{\textstyle\frac{1}{2}}\ave{\hat I_\mathrm{sin}^{(1)}\hat I_\mathrm{sin}^{(2)}}\,, \\
\label{eq:Cquaddef}
\mathrm{Im}\{\ave{\hat I_\Omega^{(1)}\hat I_{-\Omega}^{(2)}}\}&={\textstyle\frac{1}{2}}
\ave{\hat I_\mathrm{sin}^{(1)}\hat I_\mathrm{cos}^{(2)}}-{\textstyle\frac{1}{2}}\ave{\hat I_\mathrm{cos}^{(1)}\hat I_\mathrm{sin}^{(2)}}.
\end{align}

Two-beam photocurrent correlation is usually obtained by considering electronic photocurrent components \textit{in phase} with one another, as given by Eq.~(\ref{eq:Cphasedef}). Direct substitution of the photocurrent operators of HD [Eq.~(\ref{eq:HDcompcossinpm})] in Eq.~(\ref{eq:Cphasedef}) shows that the real part of spectral photocurrent correlations, $\mathrm{Re}\{\ave{\hat I_\Omega^{(1)}\hat I_{-\Omega}^{(2)}}\}$, 
retrieves the two-beam correlation block $\mathbf{C}_{s}^{(12)}$ [Eq.~(\ref{eq:Vplus12Vminus12})]. Explicitly, one has
\begin{eqnarray}
\mathrm{Re}\{\ave{\hat I_\Omega^{(1)}\hat I_{-\Omega}^{(2)}}\}&=&\cos\varphi_1\cos\varphi_2\,\mu+\sin\varphi_1\sin\varphi_2\,\nu\\
& &+\cos\varphi_1\sin\varphi_2\,\xi+\sin\varphi_1\cos\varphi_2\,\zeta,\nonumber
\end{eqnarray}
where $\varphi_j$ are independently controllable phases of LOs. 
Thus, in the usual experimental procedure, HD allows the complete determination either of the symmetric or the anti-symmetric covariance matrix of two beams [Eq.~(\ref{eq:Vplus12Vminus12})]. However, to access complete four-mode information, one needs to determine  in addition the correlations between $\mathcal{A}$ and $\mathcal{S}$ modes of the two beams [Eq.~(\ref{eq:V4mode})]. For the case of a single beam, that is the point where HD fails. As we now show, the same limitation does not affect the quantum noise of two beams if a slight improvement is applied to the usual experimental setup of spectral HD. 

The correlations of two-beam $\mathcal{S}/\mathcal{A}$ modal subspaces appear in the photocurrent signal by correlating electronic components \textit{in quadrature}, i.e. as in Eq.~(\ref{eq:Cquaddef}). HD retrieves for this experimental signal 
the expression
\begin{eqnarray}
\label{eq:iomegaquad}
\mathrm{Im}\{\ave{\hat I_\Omega^{(1)}\hat I_{-\Omega}^{(2)}}\}=&\cos\varphi_1\sin\varphi_2\,\kappa+\sin\varphi_1\cos\varphi_2\,\lambda\\
&+\sin\varphi_1\sin\varphi_2\,\tau+\cos\varphi_1\cos\varphi_2\,\eta\nonumber
\end{eqnarray}
recovering all moments appearing in Eq.~(\ref{eq:C4modest}), except for the single-beam `hidden' moment $\delta^{(j)}$.

The technique of HD is indeed sensitive to a broader set of two-beam correlations if the real and imaginary parts of $ \ave{\hat I_\Omega^{(1)}\hat I_{-\Omega}^{(2)}}$ are measured together. That could be realized by improving the usual experimental apparatus of spectral HD with the addition of an eLO in quadrature with the usual one [Fig.~\ref{fig:Icossin}]. 
Differently from the single beam case, since there are two independent optical local oscillators, it is possible to vary independently the measured quadratures of $\mathcal{S}$ and $\mathcal{A}$ modes of different beams. 

In RD, using the photocurrent operators of Eq.~(\ref{eq:Jcossin}), it is straightforward to establish that
the real and imaginary parts of $\ave{\hat J_\Omega^{(1)}\hat J_{-\Omega}^{(2)}}$
 are individually sensitive to the totality of two-beam correlation moments, although with differing coefficients. Explicitly, 
\begin{eqnarray}
\label{eq:jomegaphase}
\mathrm{Re}\{\ave{\hat J_\Omega^{(1)}\hat J_{-\Omega}^{(2)}}\}= &\,c_\mu\,\mu+c_\nu\,\nu+c_\kappa\,\kappa+c_\lambda\,\lambda\\
&+c_\xi\,\xi+c_\zeta\,\zeta+c_\eta\,\eta+c_\tau\,\tau,\nonumber\\
\label{eq:jomegaquad}
\mathrm{Im}\{\ave{\hat J_\Omega^{(1)}\hat J_{-\Omega}^{(2)}}\}= &c_\eta\,\mu+c_\tau\,\nu+c_\xi\,\kappa+c_\zeta\,\lambda\\
&+c_\kappa\,\xi+c_\lambda\,\zeta+c_\mu\,\eta+c_\nu\,\tau.\nonumber
\end{eqnarray}
where $c_\mu,c_\eta,c_\nu,c_\tau,c_\xi,c_\kappa,c_\zeta,c_\lambda$ are real functions of detunings $\Delta^{(j)}$ defined, with the help of Eq.~(\ref{defineg}), by  $2g_{+}^{*(1)}g_{+}^{(2)}=c_\mu-ic_\eta$ , $2g_{-}^{*(1)}g_{-}^{(2)}=c_\nu-ic_\tau$ , $2g_{+}^{*(1)}g_{-}^{(2)}=c_\xi-ic_\kappa$ , and $2g_{-}^{*(1)}g_{+}^{(2)}=c_\zeta-ic_\lambda$.  We note that since cross correlations involve two beams, no vacuum noise contributes to correlation signals. 

Another interesting point  comes from the fact that $\mathrm{Re}\{\ave{\hat J_\Omega^{(1)}\hat J_{-\Omega}^{(2)}}\}$ and
$\mathrm{Im}\{\ave{\hat J_\Omega^{(1)}\hat J_{-\Omega}^{(2)}}\}$
 are somewhat redundant, since they depend on the same unknown moments. In reality, apart from the fact that redundancy improves experimental precision, these signals present varying sensitivity to different moments. Hence each signal is better suited to provide information about a given set of moments.

\section{Semi-classical noise picture and the spectral matrix}
\label{subsec:efectivesinglemode}

The description presented in this section is centered on the photocurrent and its understanding as a semi-classical quantity. It directly connects the complex Fourier photocurrent components with proper quadratures of field modes. When treating the evolution equations of physical systems such as the optical parametric oscillator, such description allows the addition of vacuum fluctuations originated from the field quantization to the linearized equations of non-linear intracavity processes, in a semi-classical approach~\cite{fabreleshouches,kimbleleshouches}. 

A single beam can be described in most experiments by a relatively narrow bandwidth source around a central frequency $\omega_0$, as stated in Eq.~(\ref{eq:emaisemenos}). Stationary physical processes creating the beams will produce fields for which the two-time correlation covariance matrix  $V(t,t+\tau)=\langle \vec X(t) \vec X(t+\tau)^T\rangle$ is independent of time $t$. In this case, a spectral matrix $S(\Omega)$ can be readily defined from the Fourier transform of 
$V(\tau)$. 

Borrowing methods from the semi-classical analysis of quantum noise, and motivated by the fact that $\mathcal{A}$ and $\mathcal{S}$ modes possess the same quantum statistics for stationary quantum states (apart from a local rotation of phase space), we employ a single-mode interpretation of photocurrent quantum fluctuations by \textit{imposing} on Eq.~(\ref{eq:Iomeganonhermitian}) the form 
\begin{equation}
\hat I_\Omega(\varphi)=\cos\varphi\,\hat{\mathcal{P}}_\Omega+\sin\varphi\,\hat{\mathcal{Q}}_\Omega \;,
\label{eq:Iomegasemicop}
\end{equation}
where $\hat{\mathcal{P}}_\Omega$ and $\hat{\mathcal{Q}}_\Omega$ are respectively defined as the amplitude and phase semi-classical ``quadrature'' operators. Using Eq.~(\ref{eq:Iomeganonhermitian}), it is simple to recognize the relation between these new ``quadrature'' operators and proper modal quadrature operators as
\begin{equation}
\label{eq:defquantumPomega}
%\hat{\mathcal{P}}_\Omega&=\frac{\hat p_++i\hat q_+}{\sqrt{2}}
\hat{\mathcal{P}}_\Omega={\textstyle \frac{1}{\sqrt{2}}}\left(\hat p_s+i\hat q_a\right), \quad
%\label{eq:defquantumQomega}
\hat{\mathcal{Q}}_\Omega={\textstyle \frac{1}{\sqrt{2}}}\left(\hat q_s-i\hat p_a\right).
\end{equation}
These quadratures are ``semiclassical'' in the sense that their counterparts, in terms of complex numbers, are an efficient way to describe the generation, evolution and detection of Gaussian states of light producing stationary photocurrent.

Although non-Hermitian, those  quadratures behave as effective single-mode quadrature operators when it comes to describe the spectral noise power and second-order moments in general~\cite{cavescomplexquadPRA85,cavescomplexquadmatrixPRA85,mandelScomplex,zavattaPRA2002}. One just has to follow a semi-classical prescription to correctly calculate quadrature noise power, given by Eq.~(\ref{eq:Snoisespectrum}). The amplitude and quadrature noise spectra then read as
\begin{align}
\label{eq:SPmaismenosa}
S_\mathcal{P}(\Omega)&\equiv\ave{\hat{\mathcal{P}}_\Omega\hat{\mathcal{P}}_{-\Omega}}={\textstyle \frac{1}{2}}\Delta^2\hat p_s+{\textstyle \frac{1}{2}}\Delta^2\hat q_a, \\
\label{eq:SQmaismenosa}
S_\mathcal{Q}(\Omega)&\equiv\ave{\hat{\mathcal{Q}}_\Omega\hat{\mathcal{Q}}_{-\Omega}}={\textstyle \frac{1}{2}}\Delta^2\hat p_a+{\textstyle \frac{1}{2}}\Delta^2\hat q_s, 
\end{align}
where their correspondences in terms of proper field mode operators are also included. 

In this case, it can be noted that $S_\mathcal{P}(\Omega)$ and $S_\mathcal{Q}(\Omega)$ respect an effective uncertainty relation in the form $S_\mathcal{P}(\Omega)S_\mathcal{Q}(\Omega)\geq1$, even though $[\hat{\mathcal{P}}_\Omega,\hat{\mathcal{Q}}_\Omega]=0$. Thus, \textit{as far as second-order moments are concerned, they behave as effective quadrature operators and can be effectively treated as such.} Physically, they are connected to a mixture of $\mathcal{S}/\mathcal{A}$ modal quadratures moments.

We define their correlation using the same prescription of Eqs.~(\ref{eq:SPmaismenosa}) and (\ref{eq:SQmaismenosa}), to obtain
\begin{align}
\label{eq:CPQmaismenosa}
C_\mathcal{PQ}(\Omega)&\equiv{\displaystyle\frac{1}{2}}\ave{\hat{\mathcal{P}}_\Omega\hat{\mathcal{Q}}_{-\Omega}+\hat{\mathcal{Q}}_{-\Omega}\hat{\mathcal{P}}_{\Omega}}\\
=& {\displaystyle\frac{1}{2}}\left[C(\hat p_s\hat q_s)-C(\hat p_a\hat q_a)+i\left(C(\hat p_s\hat p_a)+C(\hat q_s\hat q_a)\right)\right].\nonumber
\end{align}
The correlation is a complex number satisfying $C_\mathcal{PQ}(\Omega)=C_\mathcal{PQ}^*(-\Omega)$. Its real part is connected with single mode correlations (i.e. the moments seen by HD), while its imaginary part contains exclusive $\mathcal{S}/\mathcal{A}$ correlations (moments `hidden' to HD). 

These second-order moments can be gathered in the \textit{complex spectral matrix}, a noise representation akin to the covariance matrix, but defined in terms of photocurrent noise as a classical quantity instead of quantum moments of field observables. The spectral matrix is 
defined as $\mathbf{S}=\ave{\vec Z_\Omega\cdot\vec Z_{-\Omega}^T}$, where $\vec Z=(\hat{\mathcal{P}}_\Omega \;\hat{\mathcal{Q}}_\Omega)^T$. Explicitly, it reads as the following $2 \times 2$ Hermitian matrix
\begin{equation}
\mathbf{S}=\left(
\begin{array}{cc}
S_\mathcal{P}(\Omega) & C_\mathcal{PQ}(\Omega)\\
C_\mathcal{PQ}^*(\Omega) & S_\mathcal{Q}(\Omega)
\end{array}
\right).
\end{equation}

Using the expressions for semi-classical quadrature noise powers in terms of proper field mode operators [Eqs.~(\ref{eq:SPmaismenosa}) and (\ref{eq:CPQmaismenosa})], it is straightforward to show that the real part of the spectral matrix can be written in terms of the covariance matrices for the symmetric and anti-symmetric combinations of sidebands presented in Eq.~(\ref{eq:Vstat}), as
\begin{equation}
\label{eq:SVplusVminuspre}
\mathrm{Re}\{\mathbf{S}\}={\textstyle \frac{1}{2}}\mathbf{V}_s +{\textstyle \frac{1}{2}}\mathbf{V}_a' \;,
\end{equation}
 where $\mathbf{V}_a'$ is the covariance matrix of mode $\mathcal{A}$ including a local rotation of quadratures by $\pi/2$ ($\hat p_a'=\hat q_a$ and $\hat q_a'=-\hat p_a$).
\textit{Hence the real part of the spectral matrix does not correspond in general to a covariance matrix, but rather to the mixture of individual covariance matrices of symmetric and anti-symmetric modes}~\cite{cavesamplifiersPRA1982}. Nevertheless, it can be understood as a proper single-mode covariance matrix in case $\mathbf{V}_s=\mathbf{V}_a'$~\cite{ralphmixedPRA08}. As shown in Ref.~\cite{gaussiantobe}, such condition is fulfilled if the measured photocurrent is Gaussian and the quantum state is assumed to be Gaussian. 

The imaginary part of the spectral matrix has a simple interpretation, representing exclusive two-mode $\mathcal{S}/\mathcal{A}$ correlations. It appears in the antidiagonal of $\mathbf{S}$ and is given by the imaginary part of $C_\mathcal{PQ}(\Omega)$. 
For stationary states  written in terms of entries of Eq.~(\ref{eq:Vstat}), $\mathbf{S}$ assumes the general form~\cite{cavesamplifiersPRA1982}
\begin{equation}
\label{eq:SVplusVminus}
\mathbf{S}=\left(
\begin{array}{cc}
\alpha & \gamma + i \delta \\
\gamma - i \delta& \beta
\end{array}
\right).
\end{equation}

This matrix can not be reconciled with the single mode approximation in the most general case (i.e. quantum states possessing spectral energy imbalance $\delta\neq0$). This brings strong limitations to the interpretation of  $\mathbf{S}$ as a covariance matrix.
Only in the case where the generated state is such that the imaginary part is zero can quantum noise then be formally interpreted as a single-mode effect, and the spectral matrix satisfies all the properties of a covariance matrix. It can then be used to formally investigate the quantum state, e.g. in testing for entanglement in an effective single-mode approximation~\cite{simon,dgcz}. When this condition is not 
satisfied~\cite{prltobe}, the complex spectral matrix can not be fully reconstructed with HD and one must resort to RD to go beyond the single mode approximation and necessarily refer to two-modes to describe quantum noise.

\subsection{Extension to multiple beams}

The analysis of two-beam photocurrent correlations in the stationary regime are also simplified by the semi-classical quadratures. We define the two-beam spectral matrix as $\mathbf{S}^{(12)}=\ave{\vec Z_\Omega^{(12)}\cdot(\vec Z_{-\Omega}^{(12)})^T}$, where $\vec Z^{(12)}=(\hat{\mathcal{P}}_\Omega^{(1)}\;\hat{\mathcal{Q}}_\Omega^{(1)}\;\hat{\mathcal{P}}_\Omega^{(2)}\;\hat{\mathcal{Q}}_\Omega^{(2)})^T$. The explicit form  
\begin{equation}
\mathbf{S}^{(12)}=\left(
\begin{array}{cc}
\mathbf{S}^{(1)} & \mathbf{C_S}^{(12)} \\
\mathbf{C_S}^{(21)} & \mathbf{S}^{(2)}
\end{array}
\right)
\label{eq:S4mode}
\end{equation}
makes it direct to relate the spectral matrix with the covariance matrix of Eq.~(\ref{eq:V4mode}).

Once more, the real part of the two-beam spectral matrix contains all information usually obtained with HD. It corresponds to the covariance matrix of modes $\mathcal{S}^{(j)}$, which for stationary states fulfill $\mathbf{V}_{s}^{(12)}=\mathbf{V}_{a}'{}^{(12)}=\mathrm{Re}\{\mathbf{S}^{(12)}\}$ 
and $\mathbf{C}_s^{(12)}=\mathrm{Re}\{\mathbf{C_S}^{(12)}\}$. 

Furthermore, correlations between modes $\mathcal{S}^{(j)}$ on one side and $\mathcal{A}^{(j')}$ on the other appear in the imaginary part of $\mathbf{C_S}^{(12)}$. Analogously to the case of a single beam, these moments are connected on the level of the four-mode covariance matrix with $\mathbf{C}_{(s/a')}^{(12)}$ [Eq.~(\ref{eq:C4modest})]. 
The relation between these matrices is
\begin{equation}
\mathbf{V}_{(s/a')}=
\left(
\begin{array}{cc}
\mathrm{Re}\{\mathbf{S}^{(12)}\} & -\mathrm{Im}\{\mathbf{S}^{(12)}\} \\
\mathrm{Im}\{\mathbf{S}^{(12)}\} & \mathrm{Re}\{\mathbf{S}^{(12)}\}
\end{array}
\right).
\end{equation}

If certain constraints on the quantum state of optical sidebands are assumed or established (namely, stationarity and lack of longitudinal two-mode correlations), {\it the spectral matrix contains the same information as the covariance matrix of modes $\mathcal{S}$ or $\mathcal{A}$}. In this case, we may either use the explicit four-mode covariance matrix $\mathbf{V}_{(s/a')}$ of two beams or adopt the simplified two-mode form $\mathbf{S}$ as the effective description of the quantum state, halving the system dimension. For such states, the spectral matrix corresponds to a partial trace of either mode $\mathcal{S}$ or mode $\mathcal{A}$ in favor of the other.

\section{Experimental results}
\label{sec:experimental}

The measurement techniques described in the preceding sections can be applied to the reconstruction of the covariance matrix of  many optical systems. We concentrate here on the case of modes $\mathcal{S}$ or $\mathcal{A}$ of pump, signal, and idler beams interacting in an above-threshold optical parametric oscillator (OPO).

When pumped above the oscillation threshold, the OPO produces three entangled beams of light~\cite{prl2006,coelhoscience09},  by means of stimulated parametric down-conversion (PDC). Detection of quantum noise with photocurrent demodulation with an eLO implies exploring an effective  six-mode quantum state. 
Hidden correlations stem from asymmetries among sidebands of different beams, as shown in previous sections, and could appear due to the richer above-threshold dynamics of energy exchange among six sideband modes. While below the threshold only sideband modes of twins (signal and idler) are expected to be populated by photon pairs, above the threshold pump beam sidebands are populated by upconversion, and thereby influence twin beam sidebands~\cite{pratobe}. 

\begin{figure}[ht]
\includegraphics[width=5cm,angle=270]{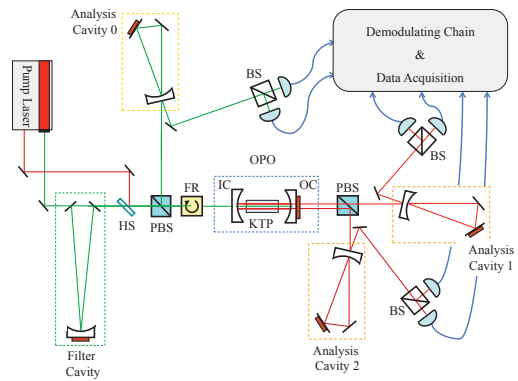}%{realimag.eps}
\caption{Setup for the reconstruction of the OPO beams' covariance matrix. PBS: polarizing beam splitter, BS: 50/50 beam splitter, HS: harmonic separator, IC: input coupler, OC: output coupler (OPO cavity), FR: Faraday rotator.}
\label{fig:VQ03}
\end{figure}

Our light source  was described in previous publications~\cite{praphononnoise,coelhoscience09}, generating three beams at the pump  (532 nm), and nondegenerate signal and idler modes (around 1064 nm) (see Fig.~\ref{fig:VQ03}). 
 The OPO consists of a type II phase-matched KTP (potassium titanyl phosphate, KTiOPO$_4$)  crystal in a linear resonator, with free spectral range of about 5~GHz and cavity finesses of  16, 135, and 115 for pump, signal, and idler modes, respectively. It is pumped by a doubled Nd:YAG laser, and has a threshold power of 67 mW. In the present measurements, the pump power was fixed at $110$ mW. We measure the output quantum states of pump, signal, and idler beams using RD with three dedicated resonators. They have nearly the same resonance bandwidth of 12(1)~MHz, and similar values of the impedance matching parameter $d\approx0.85$. This configuration enables the detection of single-beam `hidden' moments although it is not optimized. 

After reflection by its respective resonator, each beam is measured with a pair of amplified photodetectors (30~MHz response bandwidth) to allow shot noise calibration by subtraction of their photocurrent signals. Quantum properties of each beam are measured by summing each pair of photocurrents. We utilize the improved technique of Fig.~\ref{fig:Icossin} to correlate photocurrent electronic quadratures. Photocurrent signals are independently mixed with two electronic references (eLO) at 21~MHz dephased by $\pi/2$, corresponding to measurement operators $\hat I^{(j)}_\theta$ and $\hat I^{(j)}_{\theta+\frac{\pi}{2}}$ (where $j=0,1,2$, respectively, denotes pump, signal, and idler beams) of Eq.~(\ref{eq:imix}). The result is filtered with 600~kHz low-pass bandwidth with the aid of a computer A/D converter card, representing a single quantum measurement (corresponding to a measurement time of $1.67 \mu s$). During state reconstruction, resonators are scanned nearly synchronously across resonance with their respective beams, and data points are registered for 450 different values of detuning. Each quantum measurement is repeated 1000 times, over which state averages are calculated and operator moments determined. Given the laser bandwidth of 1~kHz, the time required for the acquisition of quantum statistics is larger than the typical time scale of phase diffusion. Our measurements are thus in the phase mixing regime. The entire procedure yields 450,000 quantum measurements in 750~ms per beam, i.e., 450 operator moments per scan as functions of detuning.

\subsection{Single-beam `hidden' correlations}

We start our experimental analysis by considering the quantum states of individual beams. We verify that measured photocurrents obey the conditions of Eq.~(\ref{eq:Iphasemix}), consistent with the phase mixing regime, as expected. Gaussianity of the photocurrent fluctuations indicate the stationarity of the quantum state within experimental precision~\cite{gaussiantobe}. 

\begin{figure}[ht]
\includegraphics[width=0.9\linewidth]{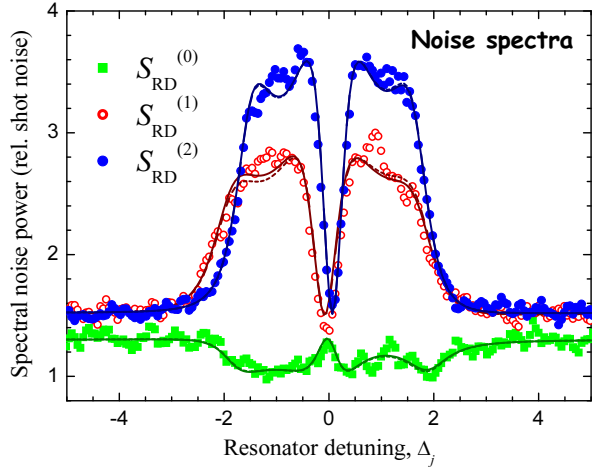}%{HF.eps}
\caption{(Color online) Pump (green squares), signal (red open circles) and idler (blue full circles) photocurrent noise power $S_\mathrm{HD}^{(j)}$ as functions of respective resonator detuning $\Delta_j$. Lines represent theoretical fits of Eq.~(\ref{eq:Srd}) either considering (solid) or disregarding (dashed) the respective `hidden' moment $\delta^{(j)}$. }
\label{fig:HF}
\end{figure}

Single beam quadrature operator moments are measured from the spectral noise power $S_\mathrm{HD}^{(j)}$ of individual photocurrents as each optical resonator is scanned across resonance. Data are presented in Fig.~\ref{fig:HF}. Three data sets refer to pump, signal, and idler photocurrent spectral noise powers. We use resonator detection to investigate the `hidden' moment $\delta^{(j)}$ representing energy imbalance between sidebands of a single beam. Solid lines represent fits of Eq.~(\ref{eq:Srd}) to the quantum noise of each beam, while dashed lines provide fits of the same equation imposing $\delta^{(j)}=0$. 

Comparison of solid and dashed lines shows that $\delta^{(j)}$ does not influence data fits within experimental precision, and is hence compatible with zero for all individual beams. 
According to the theoretical model describing the OPO, energy imbalance between sidebands of a single beam is not expected, since the bandwidth of the PDC process is many orders of magnitude larger than their frequency difference, and the OPO is operated on triple resonance. Measurements hence agree with theoretical expectations. 
In this case, resonator detection provides experimental support to the effective single-mode approximation in the treatment of individual quantum states of pump, signal and idler beams.

\subsection{Two-beam `hidden' correlations}

We now investigate all possible two-beam correlations that would be missed by the usual homodyne detection.
\begin{figure}[ht]
\includegraphics[width=0.9\linewidth]{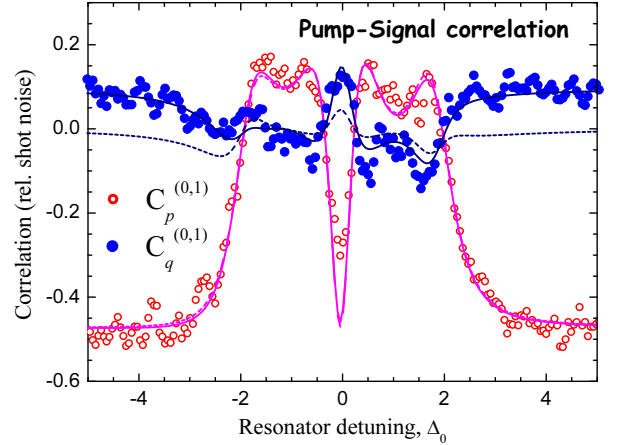}%{VQ01.eps}
\caption{(Color online) Quantum correlations between photocurrent components of different beams as functions of resonator detuning. Correlations in phase $C^{(12)}_\mathrm{p}$ (red open circles) and in quadrature $C^{(12)}_\mathrm{q}$ (blue solid circles) are depicted. Theoretical fits of Eqs.~(\ref{eq:jomegaphase}) and~(\ref{eq:jomegaquad}) to the data either consider (solid lines) or disregard (dashed lines) `hidden' four-mode correlations. }
\label{fig:VQ01}
\end{figure}
Data of pump--signal correlations in phase $C^{(01)}_\mathrm{p,RD}\equiv\mathrm{Re}\{\ave{\hat I_\Omega^{(0)}\hat I_{-\Omega}^{(1)}}\}$ and in quadrature $C^{(01)}_\mathrm{q,RD}\equiv\mathrm{Im}\{\ave{\hat I_\Omega^{(0)}\hat I_{-\Omega}^{(1)}}\}$ are presented in Fig.~\ref{fig:VQ01} as resonators are scanned in near synchrony. 
Theoretical fits of Eqs.~(\ref{eq:jomegaphase}) and (\ref{eq:jomegaquad}) determine the best correlation matrix of Eq.~(\ref{eq:C4modest}) to fully account for all data sets together, i.e. $C^{(01)}_\mathrm{p,RD}$, $C^{(01)}_\mathrm{q,RD}$ and individual power spectra $S_\mathrm{RD}^{(0)}$ and $S_\mathrm{RD}^{(1)}$. Coefficients $c_\mu,c_\eta,c_\nu,c_\tau,c_\xi,c_\kappa,c_\zeta,c_\lambda$ are determined independently as functions of $\Delta_0$ and $\Delta_1$ by monitoring LO power reflected across resonances. 

Two types of data fit are calculated to help isolate the influence of `hidden' moments. Solid lines result from fits of the most general stationary quantum state of Eq.~(\ref{eq:C4modest}). Dashed lines impose `hidden' moments as null, i.e. $\kappa=\tau=\eta=\lambda=0$. 

As seen in Fig.~\ref{fig:VQ01}, the photocurrent correlation in phase $C^{(01)}_\mathrm{p,RD}$ is not very sensitive to `hidden' quadrature operator moments, since its features are well accounted for by both solid and dashed line curves: `Hidden' moments do not need to be invoked to explain $C^{(01)}_\mathrm{p,RD}$. The contribution of these `hidden' moments to Eq.~(\ref{eq:jomegaphase}) is thus small in comparison to the contributions of other moments for our particular quantum state, given the nearly synchronous scanning of analysis cavities. The scenario is inverted in the data for correlations in quadrature $C^{(01)}_\mathrm{q,RD}$. Now large deviations can be observed by comparing the two types of theoretical fits, rendering `hidden' moments essential to explain the measurements. 

\begin{figure}[ht]
\includegraphics[width=0.9\linewidth]{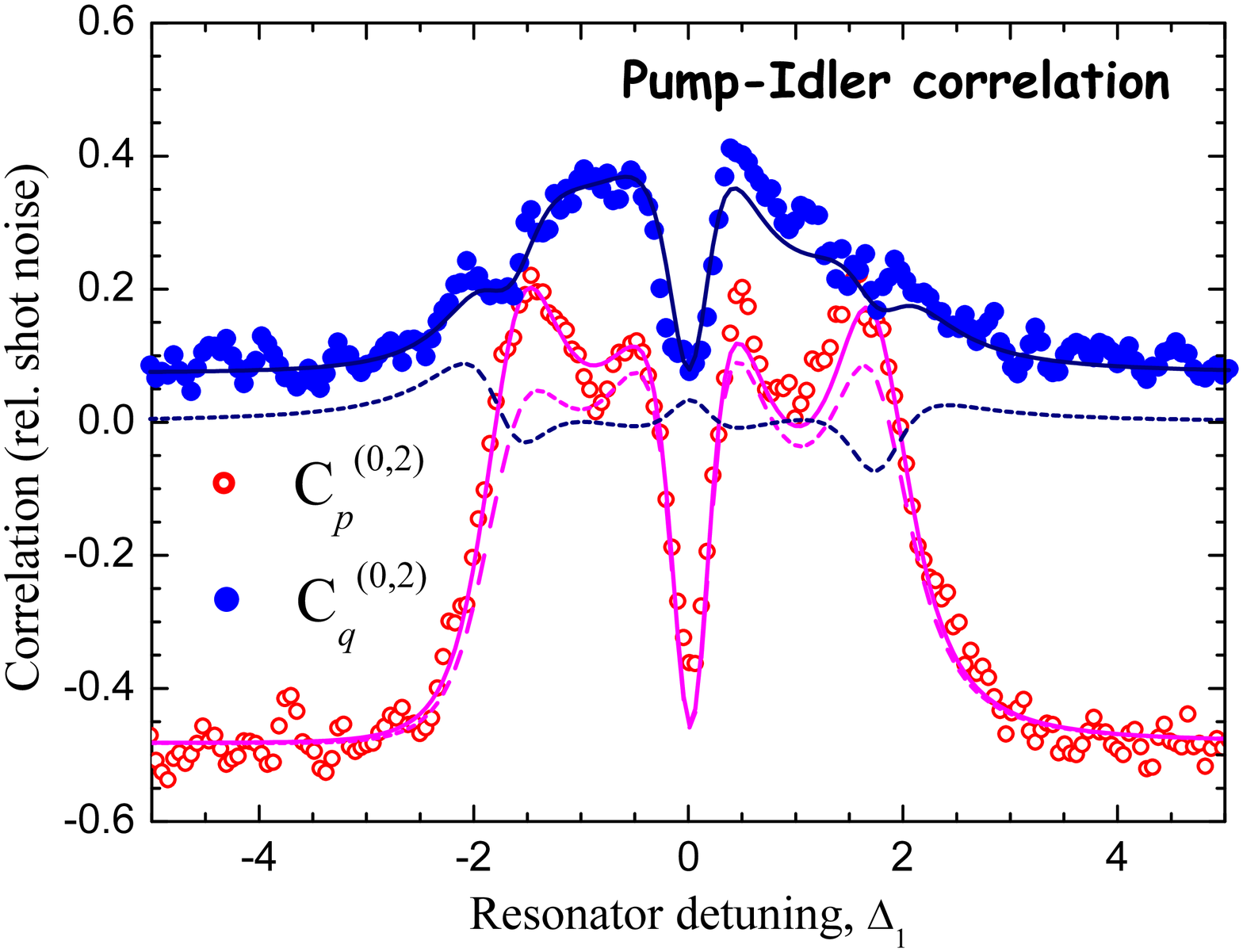}%{VQ02.eps}
\caption{(Color online) 
Same as Fig. \ref{fig:VQ01}, for pump and idler modes.}
\label{fig:VQ02}
\end{figure}

Similar results are shown in Fig.~\ref{fig:VQ02} concerning pump--idler beams. In this case, the photocurrent correlation in phase $C^{(02)}_\mathrm{p,RD}$ presents better sensitivity to the presence of `hidden' moments, although not sufficient to produce quantitative results. Their existence is once more better established by the correlations in quadrature $C^{(02)}_\mathrm{q,RD}$, for which stronger deviations between theoretical fits of solid and dashed lines can be seen. Owing to experimental asymmetries between twin beam beams, such as imbalanced signal-idler losses inside the OPO resonator, pump--idler beams show stronger `hidden' correlations than pump-signal beams.

\begin{figure}[ht]
\includegraphics[width=0.95\linewidth]{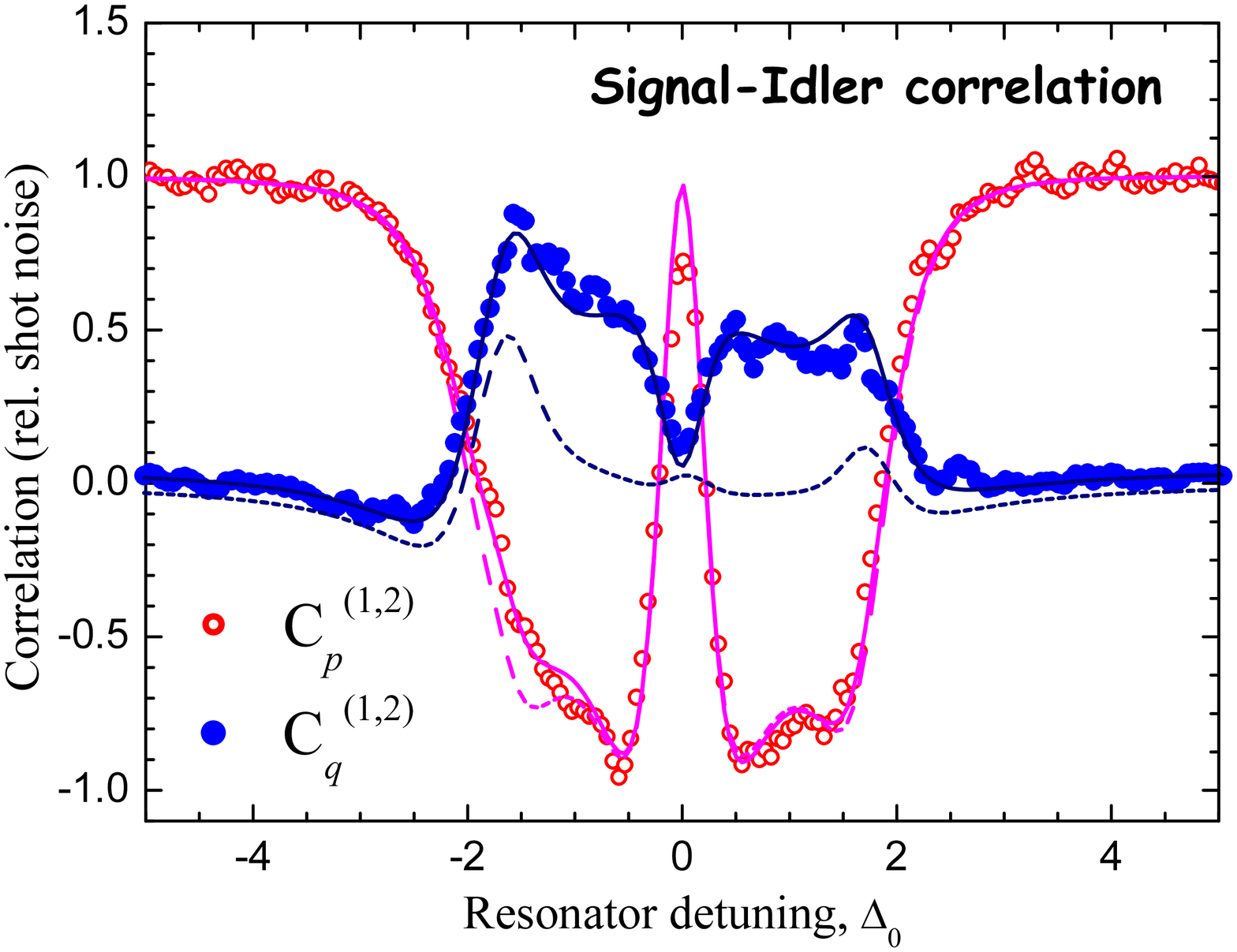}%{realimag.eps}
\caption{(Color online) 
Same as Fig. \ref{fig:VQ01}, for pump and idler modes.}
\label{fig:VQ12}
\end{figure}

Concluding the complete characterization of the OPO, we proceed with the analysis of signal and idler correlations in the same way in Fig.~\ref{fig:VQ12}, where the same conclusions apply. 
Given the Gaussianity tests we have applied to the OPO \cite{gaussiantobe}, we are left with the single assumption of stationarity of the process to claim that we have performed a complete measurement of the six mode covariance matrix of the OPO, at the analysis frequency of 21~MHz. A complete description of the hexapartite mode produced in this system, including all the terms that are unreachable by the usual homodyning detection is thus possible. The measured spectral matrix, with entries normalized to the standard quantum level(SQL), has the following real and imaginary parts:

\begin{eqnarray}
\mathrm{Re}&&\{\mathbf{S}^{(012)}\}=\mathbf{V}^{(012)}_s\\ \nonumber
&& \left(
\begin{array}{cccccc}
1.30& -0.07 & -0.47& 0.00&-0.48&-0.03\\
		&1.07 &0.12	&0.16&0.14&0.08\\
		&			&1.52 &-0.02&1.00&0.05\\
		&			&			&2.87 &0.05&-0.91\\
		&			&			&			&1.52 &-0.05\\
		&			&			&			&			& 3.64
\end{array}
\right)
\label{eq:Sreal}
\end{eqnarray}

\begin{eqnarray}
\mathrm{Im}&&\{\mathbf{S}^{(012)}\}=\mathbf{C}^{(012)}_{(s/a')}\\ \nonumber
& &\left(
\begin{array}{cccccc}
0 	&   \mathbf{-0.04}	& 0.10	& 0.04	&0.07	&0.14	\\
	&0 	&-0.03	&-0.03	&-0.02	&0.38	\\
	&	&0 	&\mathbf{0.34}	&0.05	&	-0.08\\
	&	&	&0 	&	0.04&	0.54\\
	&	&	&	&0 	&\mathbf{0.17}	\\
	&	&	&	&	& 0
\end{array}
\right)
\label{eq:SImag}
\end{eqnarray}

In most cases the $\delta$ term (bold in the matrix) is small compared to the other terms of the matrix. It is compatible with zero for signal and idler modes given the uncertainty of $0.2$ SQL for these terms (this is a relatively large value for the uncertainty, which can be improved by optimizing the analysis cavities for the detection of the $\delta$ term). Considering the narrow bandwidth of the OPO cavity, a nonzero value can originate from small deviations from the exact resonance condition during the OPO operation. Such terms are inaccessible with the homodyne detection technique, but can be observed with resonator detection. Although their effect is nearly negligible in our data (as observed in Fig.~\ref{fig:HF}), it is important to experimentally determine them as such, in order to achieve complete reconstruction of an unknown quantum state~\cite{prltobe}. 

Cross correlations between symmetric and anti-symmetric modes of different fields are clearly present beyond the overall uncertainty, which is below $0.05$ SQL for these terms. We observe 
larger values for the $\lambda$ correlations between the phase quadratures. This is not surprising, given other sources of phase noise such as phonons~\cite{praphononnoise}. 
Nonzero values are also observed for $\kappa$ and $\eta$, with $\tau$ compatible with zero [see Eq.~\ref{eq:C4modest}]. The measurement of these parameters could also be obtained from homodyne detection, although only by employing the improved quadrature demodulation scheme of Fig~\ref{fig:Icossin}, as in our current setup. The nonzero values demonstrate that information about the quantum state exists in all six-modes and that by reducing the system dimension to an effective three-mode description quantum state information is lost, inducing an artificial loss of purity and possibly of quantum properties such as entanglement.

\section{Concluding Remarks}
\label{sec:conclusao}

The quantum noise of light beams is an inherent multimode effect, even for a single beam. Each spectral component of the measured
quantum noise has information on the collective quantum state of two optical field modes. 

Spectral noise power as measured with homodyne detection (HD) provides insufficient information to reconstruct the two-mode Gaussian quantum state of a single beam without prior knowledge. 
We have shown, on the other hand, that the alternative technique of resonator detection (RD)
allows the contributions of individual sideband modes to be identified in the spectrum of quantum noise, by providing modal dependent loss and phase shifts by means of manipulation of an optical resonance. By retrieving quantum state information beyond the single-mode approximation of HD, the technique allows the single-mode approximation to be verified or discarded in the experiment. The technique enables the full reconstruction of unknown collective quantum states of the field in the ideal case of phase locked detection, also in the case of multiple beams of light. 

In most experiments, phase diffusion between the optical and electronic local oscillators leads to an inherent mixture of measurement operators. 
In this case, even additional experimental evidence, such as stationarity of the photocurrent electronic quadrature components, does not provide the amount of information needed to characterize the complete state, imposing an effective single-mode approximation to the description of quantum noise. Although such an approximation may be valid in most experiments, it remains a tacit assumption and must be recognized as such. Even in this situation, we have shown that RD is able to recover more information on the quantum state than HD, namely the energy asymmetry between sidebands~\cite{prltobe}. 

Both techniques are very similar regarding measurement efficiency. Apart from photodetector efficiency, which is a common limitation for both techniques, HD efficiency is mainly limited by the spatial overlap between LO mode and the `dim' quantum modes of interest. In a similar manner, RD finds its limitation in efficiency mainly caused by imperfect mode coupling with the resonator. In both cases, very high efficiencies ($>99\%$) are routinely attained in experiments and do not represent a distinguishing factor between techniques. 

We successfully applied resonator detection to the complete reconstruction of the quantum state, assumed to be Gaussian, of sideband modes produced by an OPO operating above the threshold.  We are now beginning to explore higher orders of multipartite entanglement~\cite{pratobe}, with implications for quantum information protocols using the continuous variables of spectral modes.

Pure quadrature operator measurements are nevertheless attainable if optical and electronic references are phase-locked to each other. The mixedness of operators used in present experiments implementing quantum information protocols in continuous variables with spectral modes, such as quantum teleportation and entanglement  swapping~\cite{teleport1,vanloockteleportwrongPRA00,praralphcriticateleport}, implies the need for assumptions regarding the quantum states. Pure operator measurements, together with the resonator detection technique analyzed here, pave the way for the implementation of unconditional quantum information protocols on completely unknown quantum states.

 \acknowledgments

We thank Y. Golubev and T. Golubeva for stimulating discussions. 
This work was supported by grants \#2010/52282-1,
\#2010/08448-2, \#2009/52157-5, S˜ao Paulo Research
Foundation (FAPESP), CNPq, CAPES (through program
PROCAD), and CNRS. CF is a member of the
Institut Universitaire de France.
This research was performed within the Brazilian National Institute 
for Science and Technology in Quantum Information (INCT-IQ) initiative.


\begin{thebibliography}{0}%
\makeatletter
\providecommand \@ifxundefined [1]{%
 \@ifx{#1\undefined}
}%
\providecommand \@ifnum [1]{%
 \ifnum #1\expandafter \@firstoftwo
 \else \expandafter \@secondoftwo
 \fi
}%
\providecommand \@ifx [1]{%
 \ifx #1\expandafter \@firstoftwo
 \else \expandafter \@secondoftwo
 \fi
}%
\providecommand \natexlab [1]{#1}%
\providecommand \enquote  [1]{``#1''}%
\providecommand \bibnamefont  [1]{#1}%
\providecommand \bibfnamefont [1]{#1}%
\providecommand \citenamefont [1]{#1}%
\providecommand \href@noop [0]{\@secondoftwo}%
\providecommand \href [0]{\begingroup \@sanitize@url \@href}%
\providecommand \@href[1]{\@@startlink{#1}\@@href}%
\providecommand \@@href[1]{\endgroup#1\@@endlink}%
\providecommand \@sanitize@url [0]{\catcode `\\12\catcode `\$12\catcode
  `\&12\catcode `\#12\catcode `\^12\catcode `\_12\catcode `\%12\relax}%
\providecommand \@@startlink[1]{}%
\providecommand \@@endlink[0]{}%
\providecommand \url  [0]{\begingroup\@sanitize@url \@url }%
\providecommand \@url [1]{\endgroup\@href {#1}{\urlprefix }}%
\providecommand \urlprefix  [0]{URL }%
\providecommand \Eprint [0]{\href }%
\providecommand \doibase [0]{http://dx.doi.org/}%
\providecommand \selectlanguage [0]{\@gobble}%
\providecommand \bibinfo  [0]{\@secondoftwo}%
\providecommand \bibfield  [0]{\@secondoftwo}%
\providecommand \translation [1]{[#1]}%
\providecommand \BibitemOpen [0]{}%
\providecommand \bibitemStop [0]{}%
\providecommand \bibitemNoStop [0]{.\EOS\space}%
\providecommand \EOS [0]{\spacefactor3000\relax}%
\providecommand \BibitemShut  [1]{\csname bibitem#1\endcsname}%
\let\auto@bib@innerbib\@empty
%</preamble>
\end{thebibliography}%


\begin{thebibliography}{99}

\bibitem{LIGO}
The LIGO Scientific Collaboration, 
%A gravitational wave observatory operating beyond the quantum shot-noise limit
Nat. Phys. \textbf{7}, 962 (2011).%, DOI: 10.1038/NPHYS2083
 
\bibitem{teleport1}
A. Furusawa \textit{et al.},
%``Unconditional Quantum Teleportation'',
Science \textbf{282}, 706 (1998).%; DOI: 10.1126/science.282.5389.706

\bibitem{teleport2}
%Teleportation of Nonclassical Wave Packets of Light
Noriyuki Lee \textit{et al.}, Science \textbf{332}, 330 (2011).%; DOI: 10.1126/science.1201034

\bibitem{storeQI}
A. E. Kozhekin, K. Molmer, and E. Polzik, 
Phys. Rev. A \textbf{62}, 033809 (2000) 
%Quantum memory for light

\bibitem{morganmitchell}
M. Napolitano \textit{et al.}, 
%M. Napolitano, M. Koschorreck, B. Dubost, N. Behbood, R. J. Sewell, M. W. Mitchell,
%Interaction-based quantum metrology showing scaling beyond the Heisenberg limit,
Nature \textbf{471}, 486 (2011). 

\bibitem{naturephotrev}
T. C. Ralph and P. K. Lam,
Nature Photon. \textbf{3}, 671 (2009).

\bibitem{squeezing}
R. E. Slusher, L. W. Hollberg, B. Yurke, J. C. Mertz, and J. F. Valley,
%Observation of Squeezed States Generated by Four-Wave Mixing in an Optical Cavity 
Phys. Rev. Lett. 55, 2409 (1985).

\bibitem{twin beams}
 A. Heidmann, R. J. Horowicz, S. Reynaud, and E. Giacobino, 
C. Fabre, and G. Camy, 
%``Observation of Quantum Noise Reduction on Twin Laser Beams'',
Phys. Rev. Lett. \textbf{59}, 2555 (1987).

\bibitem{entangledkimble}
Z. Y. Ou, S. F. Pereira, H. J. Kimble, and K. C. Peng, 
%Realization of the Einstein-Podolsky-Rosen paradox for continuous variables
Phys. Rev. Lett. \textbf{68}, 3663 (1992).

\bibitem{raymer}
D. T. Smithey, M. Beck, M. G. Raymer, and A. Faridani,
Phys. Rev. Lett. \textbf{70}, 1244 (1993).

\bibitem{raymerrmp}
A. I. Lvovsky and M. G. Raymer, 
``Continuous-variable optical quantum-state tomography'',
Rev. Mod. Phys. \textbf{81}, 299 (2009).

\bibitem{yurkemedidasqzPRA85}
B. Yurke,
%``Squeezed-coherent-state generation,''
Phys. Rev. A {\bf 32}, 300 (1985).

\bibitem{prltobe}
F. A. S. Barbosa \textit{et al.}, 
% F. A. S. Barbosa, A. S. Coelho, K. N. Cassemiro, P. Nussenzveig, C. Fabre, M. Martinelli, and  A. S. Villar
% ``Beyond spectral homodyne detection: complete quantum measurement of spectral modes of light,''
arXiv:1308.5650 [quant-ph].

\bibitem{levenson}
G. C. Bjorklund {\it et al.}, 
%G. C. Bjorklund, M. D. Levenson, W. Lenth, and C. Ortiz,
Appl. Phys. B {\bf 32}, 145 (1983).

\bibitem{shelby}
R. M. Shelby, M. D. Levenson, S. H. Perlmutter, R. G. DeVoe, and D. F. Walls,
%Broad-Band Parametric Deamplification of Quantum Noise in an Optical Fiber
Phys. Rev. Lett. \textbf{57}, 691 (1986).

\bibitem{optcommunLvovsky}
R. Kumar, E. Barrios, A. MacRae, E. Cairns, E. H. Huntington, and A. I. Lvovsky,
%``Versatile wideband balanced detector for quantum optical homodyne tomography''
Opt. Commun. \textbf{285}, 5259 (2012). 

\bibitem{zhangPRA2003}
Jing Zhang, 
%Einstein-Podolsky-Rosen sideband entanglement in broadband squeezed light
Phys. Rev. A \textbf{67}, 054302 (2003).

\bibitem{huntingtonPRA05}
E. H. Huntington {\it et al.},
%``Demonstration of the spatial separation of the entangled quantum sidebands of an optical field,''
Phys. Rev. A {\bf 71}, 041802(R) (2005).

\bibitem{glauber}
R. J. Glauber, 
Phys. Rev.\textbf{130}, 2529 (1963).

\bibitem{yurkewidebandPRA85}
B. Yurke,
%``Wideband photon counting and homodyne detection,''
Phys. Rev. A {\bf 32}, 311 (1985).

\bibitem{optcommEletQuad}
A. Heidmann, S. Reynaud, and C. Cohen-Tannoudji, 
Opt. Commun. \textbf{52}, 235 (1984).

\bibitem{shapirocossin}
H. P. Yuen and J. H. Shapiro, 
%Optical communication with ... - part III 
IEEE Trans. Inf. Theory {\bf 26}, 78 (1980). 

\bibitem{cavesamplifiersPRA1982}
C. M. Caves,
%``Quantum limits on noise in linear amplifiers'',
Phys. Rev. D {\bf 26}, 1817 (1982).

\bibitem{ralphmixedPRA08}
T. C. Ralph, E. H. Huntington, and T. Symul,
%``Single-photon side bands,''
Phys. Rev. A {\bf 77}, 063817 (2008).

\bibitem{shapirohd1}
H. P. Yuen and J. H. Shapiro, 
IEEE Trans. Inf. Theory  {\bf 24}, 657 (1978).

\bibitem{shapirohd2}
J. H. Shapiro, H. P. Yuen, and J. A. Machado Mata, 
IEEE Trans . Inf. Theory, {\bf 25}, 179 (1979).

\bibitem{yuenchanOL1983}
H. P. Yuen and V. W. S. Chan, Opt. Lett. {\bf 8}, 177 (1983).

\bibitem{optlettschum}
B. L. Schumaker, Opt. Lett. \textbf{9}, 189 (1984).

\bibitem{shapirorev1985}
J. H. Shapiro, IEEE J. Quantum Elect. \textbf{21}, 237 (1985).

\bibitem{jmo1987}
M. J. Collett, R. Loudon and C.W. Gardiner,
J. Mod. Opt. \textbf{34}, 881 (1987).

\bibitem{coherencefiction}
K. M\o{}lmer, 
%``Optical coherence: A convenient fiction,''
Phys. Rev. A \textbf{55}, 3195 (1997).
 
%\bibitem{qtomography}
%K. Vogel and H. Risken,
%Phys. Rev. A \textbf{40}, 2847 (1989).
%Determination of quasiprobability distributions in terms of probability distributions for the rotated quadrature phase

\bibitem{simon}
R. Simon,
Phys. Rev. Lett. {\bf 84}, 2726 (2000).

\bibitem{dgcz}
L. M. Duan, G. Giedke, J. I. Cirac, and P. Zoller,
%L. M. Duan {\it et al.}, 
Phys. Rev. Lett. {\bf 84}, 2722 (2000).
%; R. Simon, ibid. {\bf 84}, 2726 (2000)

\bibitem{huntington2002}
E.H. Huntington and T.C. Ralph, 
J. Opt. B: Quantum Semiclassical Opt. \textbf{4}, 123 (2002).

\bibitem{galatola}
P. Galatola {\it et al.},
Opt. Commun. {\bf 85}, 95 (1991).

\bibitem{zhangJOSAB2000}
Jing Zhang \textit{et al.}, 
J. Opt. Soc. Am. B \textbf{17}, 1920 (2000). 

\bibitem{zavattaPRA2002}
A. Zavatta, F. Marin, and G. Giacomelli,
Phys. Rev. A \textbf{66}, 043805 (2002).

\bibitem{villarajp}
A. S. Villar, 
Am. J. Phys. {\bf 76}, 922 (2008).

\bibitem{naturebreitenbach}
G. Breitenbach, S. Schiller, J. Mlynek, 
Nature \textbf{387}, 471 (1997).

\bibitem{phaselockingPRA10}
T. Kawakubo and K. Yamamoto,
%``Optical homodyne detection in view of the joint probability distribution,''
Phys. Rev. A {\bf 82}, 032102 (2010).

\bibitem{gaussiantobe}
A. S. Coelho \textit{et al.}, in preparation.

\bibitem{fabreleshouches}
C. Fabre, S. Reynaud, 
%``Quantum noise of optical fields : a semi-classical approach'', 
Les Houches Session 53, p. 679, J. Dalibard and J. M. Raymond, eds. (Elsevier Science Publisher BV, 1992).

\bibitem{kimbleleshouches}
H. J. Kimble, Les Houches Session 56, p. 603, J. Dalibard and J. M. Raimond, eds. (North Holland, Amsterdam, 1992).

\bibitem{cavescomplexquadPRA85}
C. M. Caves and B. L. Schumaker,
%New formalism for ... - part I
Phys. Rev. A {\bf 31} 3068 (1985).

\bibitem{cavescomplexquadmatrixPRA85}
B. L. Schumaker and C. M. Caves,
%New formalism for ... - part II
Phys. Rev. A {\bf 31} 3093 (1985).

\bibitem{mandelScomplex}
Z. Y. Ou, C. K. Hong, L. Mandel,
J. Opt. Soc. Am. B \textbf{4}, 1574 (1987).

\bibitem{prl2006}
A. S. Villar, M. Martinelli, C. Fabre, and P. Nussenzveig,
%A. S. Villar {\it et al.},
Phys. Rev. Lett. {\bf 97}, 140504 (2006).

\bibitem{coelhoscience09}
A. S. Coelho {\it et al.},
%A. S. Coelho, F. A. S. Barbosa, K. N. Cassemiro, A. S. Villar, M. Martinelli, and P. Nussenzveig,
%``Three-Color Entanglement"
Science \textbf{326}, 823 (2009).

\bibitem{pratobe}
F. A. S. Barbosa \textit{et al.}, in preparation.

\bibitem{praphononnoise}
 J. E. C\'esar \textit{et al.}, 
 Phys. Rev. A, \textbf{79}, 063816 (2009).

\bibitem{vanloockteleportwrongPRA00}
P. van Loock, S. L. Braunstein, and H. J. Kimble,
%``Broadband teleportation,''
Phys. Rev. A {\bf 62}, 022309 (2000).

\bibitem{praralphcriticateleport}
J. G. Webb, T. C. Ralph, and E. H. Huntington,
Phys. Rev. A {\bf 73}, 033808 (2006).

%%%%%%%%%%%%%%%%%%%%%%%%%%%%%%%%%%%%%%


%\bibitem{LOphasePRA89}
%B. Huttner and Y. Ben-Aryeh,
%%``Homodyne and heterodyne detection of squeezed vacuum states,''
%Phys. Rev. A {\bf 40}, 2479 (1989).

%\bibitem{raymerJOSAB95}
%M. G. Raymer, J. Cooper, H. J. Carmichael, M. Beck, and D. T. Smithey,
%%%``Ultrafast measurement of optical-field statistics by dc-balanced homodyne detection,'' 
%J. Opt. Soc. Am. B {\bf 12}, 1801 (1995).

%\bibitem{simonmukunda}
% R. Simon, E. C. G. Sudarshan, and N. Mukunda, Phys. Rev.
%A \textbf{36}, 3868 (1987); R. Simon, N. Mukunda, and B. Dutta,
%Phys. Rev. A \textbf{49}, 1567 (1994).


 \end{thebibliography}
\end{document}